\documentclass[fleqn,usenatbib]{mnras}

\usepackage{newtxtext,newtxmath}

\usepackage[T1]{fontenc}

\DeclareRobustCommand{\VAN}[3]{#2}
\let\VANthebibliography\thebibliography
\def\thebibliography{\DeclareRobustCommand{\VAN}[3]{##3}\VANthebibliography}

\usepackage{graphicx}	
\usepackage{amsmath}	
\usepackage{newtxtext,newtxmath}
\usepackage{relsize}
\usepackage{wrapfig, subcaption, booktabs, longtable, caption, makecell}
\usepackage{orcidlink}

\newcommand{\ols}[1]{\mskip.5\thinmuskip\overline{\mskip-.5\thinmuskip {#1} \mskip-.5\thinmuskip}\mskip.5\thinmuskip} 

\newcommand\closure[1]{
  \tctestifnum{\count@stringtoks{#1}>1} 
  {\ols{#1}} 
  {\olsi{#1}} 
}

\newcommand{\lowMh}{mid\_$M_h$\:}
\newcommand{\highMh}{high\_$M_h$\:}
\newcommand{\lowLX}{low\_$L_X$\:}	
\newcommand{\highLX}{high\_$L_X$\:}
\newcommand{\lowEX}{low\_$E_0$\:}
\newcommand{\highEX}{high\_$E_0$\:}
\newcommand{\fidnoTS}{Fid\_no$T_S$\:}
\newcommand{\GENESIS}{\textsc{genesis}\:}
\newcommand{\MERAXES}{\textsc{meraxes}}
\newcommand{\fastcm}{\textsc{21cmfast}\:}

\title[21cm Skew Spectrum]{The Impact of ionization Morphology and X-ray Heating on the Cosmological 21cm Skew Spectrum}

\author[J.~H. Cook et al.]{
J. H. Cook$^{1,2}$\thanks{E-mail: Jaiden.cook@curtin.edu.au},
S. Balu$^{2,3}$,
B. Greig$^{2,3,4}$,
C. M. Trott$^{1,2}$,
J. L. B. Line$^{1,2}$,
Y. Qin$^{2,3}$, 
J. S. B. Wyithe$^{2,4}$\\
$^{1}$International Centre for Radio Astronomy Research, Curtin University, Perth, Australia\\
$^{2}$ARC Centre of Excellence for All Sky Astrophysics in 3D (ASTRO 3D)\\
$^{3}$School of Physics, University of Melbourne, Parkville, VIC 3010, Australia\\
$^{4}$Research School of Astronomy \& Astrophysics, Australian National University, Canberra, ACT 2611, Australia
}

\date{Accepted XXX. Received YYY; in original form ZZZ}

\pubyear{2024}

\begin{document}
\label{firstpage}
\pagerange{\pageref{firstpage}--\pageref{lastpage}}
\maketitle

\begin{abstract}
The cosmological 21cm signal offers a potential probe of the early Universe and the first ionizing sources. Current experiments probe the spatially-dependent variance (Gaussianity) of the signal through the power spectrum (PS). The signal however is expected to be highly non-Gaussian due to the complex topology of reionization and X-ray heating. We investigate the non-Gaussianities of X-ray heating and reionization, by calculating the skew spectrum (SS) of the 21cm signal using \MERAXES, which couples a semi-analytic galaxy population with semi-numerical reionization simulations. The SS is the cross-spectrum of the quadratic temperature brightness field with itself. We generate a set of seven simulations from $z = 30$ to $z = 5$, varying the halo mass threshold for hosting star-formation, the X-ray luminosity per star-formation rate, and the minimum X-ray energy escaping host galaxies. We find the SS is predominantly negative as a function of redshift, transitioning to positive towards the start of reionization, and peaking during the midpoint of reionization. We do not see a negative dip in the SS during reionization, likely due to the specifics of modelling ionization sources. We normalise the SS by the PS during reionization isolating the non-Gaussianities. We find a trough ($k\sim\,0.1\,\textrm{Mpc}^{-1}$) and peak ($k\sim\,0.4-1\,\textrm{Mpc}^{-1}$) in the normalised SS during the mid to late periods of reionization. These correlate to the ionization topology, and neutral islands in the IGM. We calculate the cosmic variance of the normalised SS, and find these features are detectable in the absence of foregrounds with the SKA\_LOW.
\end{abstract}




\section{Introduction}
The cosmological 21cm neutral hydrogen line promises to be an insightful probe of the first luminous sources and the structure of the Universe during early cosmic time. The first luminous sources (stars, galaxies, compact objects) heat and ionize the surrounding intergalactic medium (IGM), through the cumulative emission of ultraviolet (UV) and X-ray photons \citep[see the following review papers: ][]{BARKANA2001,Morales2010,Pritchard_2012,Furlanetto2016}.\footnote{Reionization predominately occurs due to UV photons, with some contribution from X-ray emission (up to $\sim10$ percent \citet{Mesinger2013}).}. These ionized bubbles grow and eventually overlap, culminating in the end of reionization by redshift $\sim5.3$ \citep{Bosman2022}. These bubbles encode information about these sources onto the cosmological 21cm temperature brightness signal \citep{Bubbles_2}. These luminous sources also heat the neutral hydrogen medium through X-ray emission, which encodes additional information about these sources \citep{Pritchard2007,Furlanetto2016}. The cosmological 21cm signal is measured relative to the cosmic microwave background (CMB), and can be either in relative emission or absorption. By measuring the 21cm signal we can construct the spatial, and line of sight distributions of neutral hydrogen. This will allow for the properties of the first luminous sources to be inferred through their influence on the cosmological 21cm signal.

Most of the focus in the 21cm cosmological community has been on measuring either the one or two-point statistics of the signal. The one point statistic experiments determine the sky averaged quantities (for example the global mean temperature). For example: The Shaped Antenna measurement of the background Radio Spectrum 3 telescope \citep[SARAS3,][]{SARAS3}; the Experiment to Detect the Global EoR Signature \citep[EDGES,][]{EDGES}. The two point statistic experiments are primarily measured by radio interferometers. The current generation of radio interferometers includes the Murchison Widefield Array \citep[MWA,][]{MWA-PhaseI,MWA-PH2}; Low-Frequency Array \citep[LOFAR,][]{LOFAR}; Hydrogen Epoch of Reionization Array \citep[HERA,][]{HERA}; the New extension in Nancay upgrading LOFAR \citep[NenuFAR,][]{NenuFAR}. 

The two point statistic experiments calculate the PS of the 21cm signal which is the Fourier transform of the two point correlation function. This measures the Gaussianity or the variance of the signal as a function of comoving spatial scale. If the signal is entirely Gaussian this would capture all the information about the 21cm signal within cosmic variance\footnote{Since we cannot truly measure the ensemble average power spectrum, we can only estimate it over some volume. Each independent realisation therefore is a random sample of the true PS with some \textit{cosmic} variance.}. The signal is however expected to be highly non-Gaussian as it evolves during the Epoch of Heating (EoH) and the Epoch of reionization (EoR) \citep{Wyithe-Morales2007,Lidz_2007}. In the former the non-Gaussianities are driven by the appearance of the first luminous sources which heat the neutral IGM primarily through X-ray emission \citep{Furlanetto2006}. During this period, the strong emission from the first luminous sources drive above average temperature contrasts relative to the IGM. Eventually as X-ray heating progresses the medium saturates driving the non-Gaussianities to the matter density \citep{Wat2018}. During the latter stages of reionization the spin temperature of the neutral hydrogen is expected to be saturated $(T_S \gg T_{\rm{CMB}})$, and therefore the non-Gaussianities are largely driven by the ionization topology \citep{Hutter2019}. Analytical estimates of the characteristic size of the ionization topology (bubbles) around individual luminous sources are $\sim10\, \textrm{cMpc}$ during the late time period of reionsation \citep{Wyithe2004,Bubbles_2,Bubble_3}. However, \citet{Lin2016} showed that the characteristic size is underestimated and is closer to $\sim 20- 100\,\textrm{cMpc}$. \citet{Lin2016} and \citet{Giri2017} demonstrate the difficulty of determining the characteristic size from the complex 3D ionization topology during reionization.

Non-Gaussianity has been shown to be important in constraining the 21cm signal, in particular during reionization, and could be important for confirming a detection of the 21cm signal \citep{Shim_2017}. Non-Gaussianity has primarily been investigated by calculating the expected 21cm bispectrum. The bispectrum is the Fourier transform of the three point correlation function \citep{peebles}, like the PS it probes the central third order moment as a function of spatial scales. The bispectrum offers a complementary picture of the 21cm signal especially during the EoR \citep{Bispec2005,Maj2018}. \citet{Wat2018} investigated the non-Gaussianity due to X-ray heating from stellar sources and high mass X-ray binaries. \citet{Hutter2019} investigated the ionization morphology and the effects on the non-Gaussianity of the 21cm signal during reionization. Numerous studies have been conducted on the bispectrum and its sensitivity during reionization \citep{Bispec2005,Shintaro_2015,Bispec_shim2016,Shim_2017,Wat2017,Maj2018,Bispec-Mondal,Majumdar2020}. \citet{Trott_bispec2019} measure the bispectrum of MWA data, looking at a gridded and non-gridded estimator, establishing upper limits. \citet{Wat2020} looked at the expected foreground bispectrum, commenting on the detectability of the 21cm bispectrum in the presence of foreground systematics. More recently \citet{Tiwari_2022} showed that the bispectrum can help constrain reionization parameters.

The bispectrum has low signal to noise relative to the PS, and is computationally intensive to measure, even with the fast Fourier transform method of \citet{Wat2017}. To mitigate the difficulties related to computation and sensitivity, much focus has been on calculating the equilateral bispectrum \citep[for example:][]{Bispec2005,Shintaro_2015,Wat2018}, as well as the squeezed bispectrum, which compresses one of the triangle mode sides \citep{Chiang_2014,Bispec-Mondal}. Other work has investigated the higher order one point statistics of the simulated 21cm signal, due to X-ray heating and Ly-$\alpha$ coupling \citep[for example:][]{Wat2015,Ross2019}. The CMB cosmology community has investigated alternatives that probe the non-Gaussianity through the cross spectrum of quadratic temperature fields with the temperature field \citep{Cooray2001}. This is called the skew spectrum (SS), and is a collapsed form of the bispectrum, compressing the information into a pseudo PS as a function of one wavenumber (Fourier modes ($k$)) \citep{Regan_2017}. Generalised in \citet{Szapudi_1997,Munshi_1998} and first used by \citet{Cooray2001}, it is now gaining interest in the 21cm community with the release of \citet[][hereafter MP23]{Ma2023} at the time of writing this paper. Again drawing on the CMB cosmology community for inspiration, \citet{Dai_2020} investigated what information can be gained by combining the PS and the SS. They found that the SS in conjunction with the PS offered increased constraints on cosmological parameters. The SS promises to have better signal to noise than the bispectrum because it integrates over all bispectrum triangle configurations for a given Fourier mode $k$. Additionally the SS can be directly compared to the PS because they can be measured at the same Fourier modes. The quadratic field cross correlation approach also makes it easy to measure the SS from simulations without having to first calculate the bispectrum.

In this work we use the updated version of \MERAXES, which couples a semi-analytic galaxy formation model with a semi-numerical reionization simulation to provide a realistic population of galaxies which can interact with the IGM through a variety of feedback effects. These feedback effects include supernovae, AGN, and photoheating, along with the infall/accretion of gas \citep{MERAXES,MERAXES-AGN,MERAXES-SNR-feedback}. \citet{Balu_box} updated \MERAXES\: to include X-ray heating and spin temperature evolution for the semi-analytic galaxy formation model. Additionally in \citet{Balu_box}, the halo merger trees for the $210\,h^{-1}\,\textrm{Mpc}$ simulations were augmented to include all atomically cooled galaxies out to $z=20$ $(\sim2\times10^7\,h^{-1}\,\textrm{M}_\odot)$. We build on the earlier work of \citet{Balu_box} by performing additional simulations varying the ionization morphology through changing the minimum mass threshold for galaxies hosting star formation, the X-ray luminosity and minimum energy threshold for X-rays escaping their host galaxies. Combined, these simulations enable the exploration of the cosmic evolution of the 21cm signal using a realistic population of galaxies from the cosmic dawn down to the completion of the EoR. These are ideal for studying the non-Gaussianity of the EoH and the EoR using the SS.

The paper is outlined as follows; in section \ref{sec:ps-ss-descritpion} we define the PS and SS. In section \ref{sec:MERAXES/21cmfast} we briefly describe the simulations performed in this work. Section \ref{sec:Stats-thermal-history} presents the thermal and ionization history of each simulation as well as the statistics as a function of redshift. Section \ref{sec:PS-SS-EoR} presents the PS, the SS and the normalised SS during the EoR for each simulation. Section \ref{sec:Detectability} discusses the detectability of the normalised SS for the future SKA\_LOW radio interferometer. We discuss and draw conclusions from the results in Section \ref{sec:Discussion-Conclusion}. The cosmology used in this work is defined by \citet{Plank2021}: $h=0.68$, $\Omega_m = 0.31$, $\Omega_b = 0.048$, $\Omega_\Lambda = 0.69$, $\sigma_8 = 0.81$, and $n_s = 0.96$. All cosmological scales are in comoving units.

\section{Power Spectrum and Skew Spectrum}\label{sec:ps-ss-descritpion}

In this section we review the PS and the SS, and how they are calculated from simulation volumes.

\subsection{Power Spectrum}\label{sec:power-spectrum}

The PS is the Fourier transform of the two point correlation function, and probes the Gaussianity of a random field as a function of spatial scale $k$ \citep{peebles}:
\begin{equation}\label{eq:PS-definition}
    \langle \Tilde{\delta}_T(\mathbf{k})\Tilde{\delta}_T(\mathbf{k^\prime}) \rangle \equiv (2\pi)^3 \delta_D(\mathbf{k} + \mathbf{k^\prime}) P(k).
\end{equation}

In Equation \ref{eq:PS-definition} the angular brackets $\langle \rangle$ denote the ensemble average over different realisations of the Universe. The Dirac delta $\delta_D(\mathbf{k} + \mathbf{k^\prime})$ restricts the average to uncorrelated modes, and $P(k)$ is the spherically averaged power spectrum. $\delta_T(\mathbf{x})$ is the brightness temperature (or density) contrast $\delta_T(\mathbf{x}) = (T(\mathbf{x})-\Bar{T}(\mathbf{x}))/\Bar{T}(\mathbf{x})$, and $\Bar{T}(\mathbf{x})$ is the mean temperature. $\Tilde{\delta}_T(\mathbf{k})$ is the three dimensional Fourier transform of $\delta_T(\mathbf{x})$, defined by\footnote{In practice we perform the Fourier transform over $\delta T_b(x)$ not $\delta_T(x)$.}:
\begin{equation}
    \Tilde{\delta}_T(\mathbf{k}) = \frac{V}{N_{\rm{pix}}}\sum \delta_T(\mathbf{x})\,e^{-i\mathbf{k}\cdot \mathbf{x}}.
\end{equation}

We define the dimensionless PS:
\begin{equation}\label{eq:PS-mK2}
    \Delta^2_{T} (k,z) \equiv \frac{k^3}{(2\pi^2) V}\, \ols{\delta T}_b^2 \langle \Tilde{\delta}_T(\mathbf{k},z)\Tilde{\delta}^*_T(\mathbf{k},z) \rangle \:\rm{mK^2},
\end{equation}
where the angular brackets now denote the incoherent average in a spherical shell of width $\Delta\log{k}=0.173$, and $\Tilde{\delta}^*_T(\mathbf{k},z)$ is the conjugate transpose of $\Tilde{\delta}_T(\mathbf{k},z)$. 

\subsection{Bispectrum and Skew Spectrum}\label{sec:skew-spectrum}

The bispectrum of the 21cm brightness temperature fluctuations is defined as:
\begin{equation}\label{eq:bispectrum}
    \langle \Tilde{\delta}_T(\mathbf{k}_1)\Tilde{\delta}_T(\mathbf{k}_2)\Tilde{\delta}_T(\mathbf{k}_3) \rangle \equiv (2\pi)^3 \delta_D(\mathbf{k}_1 + \mathbf{k}_2 + \mathbf{k}_3) B(k_1,k_2,k_3),
\end{equation}
again the angular brackets and the Dirac delta function denote the ensemble average. For the bispectrum the ensemble average is over all triplet values that satisfy the closed triangle condition $\mathbf{k}_1 + \mathbf{k}_2 + \mathbf{k}_3 = 0$. Without loss of generality we let $\mathbf{k}_1 = \mathbf{k}$, $\mathbf{k}_2 = \mathbf{q}$, and $\mathbf{k}_3 = -\left(\mathbf{k} + \mathbf{q} \right)$. The SS is the integral of the bispectrum $B(k,q,|\mathbf{k}+\mathbf{q}|)$ over all possible triangle configurations, for a fixed triangle side $k$:
\begin{equation}\label{eq:skew-spectrum}
    S_{\gamma} (k) = \frac{1}{(2\pi)^3} \int_{\mathbb{R}} d^3 q\, B(k,q,|\mathbf{k}+\mathbf{q}|).
\end{equation}

It can be shown that Equation \ref{eq:skew-spectrum} is equivalent to the cross spectrum of the mean subtracted squared temperature field, to the temperature field:
\begin{equation}\label{eq:SS-definition}
    \langle \Tilde{\delta}_{T^2}(\mathbf{k})\Tilde{\delta}_T(\mathbf{k^\prime}) \rangle \equiv (2\pi)^3 \delta_D(\mathbf{k} + \mathbf{k^\prime}) S_\gamma(k),
\end{equation}
Similarly to the PS definition in Equation \ref{eq:PS-definition} the delta function and the angular brackets denote the ensemble average. We define the Fourier transform of the squared temperature field $\Tilde{\delta}_{T^2}(\mathbf{k})$ below:
\begin{equation}
    \Tilde{\delta}_{T^2}(\mathbf{k}) = \frac{V}{N_{\rm{pix}}}\sum \left(\delta_T(\mathbf{x})\right)^2\,e^{-i\mathbf{k}\cdot \mathbf{x}}.
\end{equation}

Similar to Equation \ref{eq:PS-mK2} we can define the dimensionless SS:
\begin{equation}\label{eq:SS-mK3}
    \Delta^2_{T^2,T} (k,z) \equiv \frac{k^3}{(2\pi^2) V}\,  \ols{\delta T}_b^3 \langle \Tilde{\delta}_{T^2}(\mathbf{k},z)\Tilde{\delta}^*_{T}(\mathbf{k},z) \rangle.
\end{equation}

We calculate the SS by performing a three dimensional Fourier transform of the mean subtracted and squared temperature field. We then take the product of this with the conjugate of the Fourier transform of the temperature field, and then average in spherical shells of width $\Delta\log{k}=0.173$. For consistency and for comparison we use the same bins for calculating the SS and the PS throughout this work.

\section{MERAXES}\label{sec:MERAXES/21cmfast}

In order to simulate the cosmic 21cm signal, we use the \MERAXES\:\citep{MERAXES} semi-analytical galaxy formation and evolution model. In this section, we give a brief summary of \MERAXES\:and refer the reader to other relevant works for further details.

\subsection{A Realistic Galaxy Population}\label{sec:meraxes_gal}

We make use of the L210\_N4320 dark matter-only simulation of the \GENESIS suite of $N$-body simulations (Power et al. in prep). L210\_N4320 has $4320^3$ dark matter particles in a cubical volume of side length $L = 210 h^{-1}\,\rm{Mpc}$ achieving a halo mass resolution of $\sim 5 \times 10^8~M_\odot $. 

The halo merger trees from L210\_N4320 were further `augmented' to a halo mass resolution of $\sim 3 \times 10^7~M_\odot $, the atomic cooling limit at $z=20$, using the Monte-Carlo algorithm code \textsc{DarkForest} \citep{MERAXES_Qiu2020}. This is achieved by sampling low-mass haloes from a conditional halo mass function that is based on the extended Press-Schechter theory \citep{Bond1991,Bower1991,Lacey1993}, after modifications to match the \textit{N}-body simulations' halo mass functions (HMFs). These haloes are then `grafted' onto the L210\_N4320 merger tree in a manner such that the final augmented HMFs agree with those from high resolution \textit{N}-body simulations \citep[see Figure 2 of ][]{Balu_box}. \textsc{DarkForest} also assigns and evolves the positions of the newly added haloes using the local halo density field and the linear continuity equation \citep[see][for further details]{MERAXES_Qiu2020,Balu_box}.  We therefore effectively create an \textit{N}-body simulation that has a statistically complete galaxy population down to redshift $z=20$; we deploy \MERAXES\:on this augmented simulation.

 The goal of \MERAXES\:is to simulate the growth and evolution of galaxies during the EoR in a self-consistent manner. This is achieved through detailed and physically motivated prescriptions for varied astrophysical phenomena such as radiative cooling of gas, star formation, supernovae (SNe) and active galactic nuclei feedback, and mergers \citep{MERAXES,MERAXES-QSO,MERAXES-AGN,MERAXES-SNR-feedback}. 

For each simulation snapshot, a dark matter halo increases its baryonic mass in proportion to the universal cosmic baryonic fraction. This mass is added to a `hot gas' reservoir from where it can cool down to form a `cold gas disk'. Following a star formation prescription based on the Kennicutt-Schmidt law \citep{Kennicutt1998}, stars are created out of this cold disk when a cold mass threshold is reached. The cadence of our simulation is constructed so that the longest time-step is $\sim16$ Myr. Hence newly formed massive stars can go SNe in the same time-step and less massive stars can survive for a few snapshots. \MERAXES\:therefore, has implementations for both instantaneous and delayed SNe feedback. The primary impact of SNe is to heat up the cold gas in a galaxy. SNe therefore move a portion of the cold gas to the hot halo and in very extreme energetic cases can even remove the gas from the galaxy altogether.

The amount of stellar mass in a galaxy fixes the amount of ionizing UV and X-ray photons that it produces. Once the local environment of a galaxy is ionized, the cooling properties of the IGM are affected. \MERAXES\:couples reionization feedback and galaxy growth by self-consistently modifying the amount of gas that is accreted onto a galaxy depending on the local UV background and the local IGM ionization state. The UV escape fraction $f_{\rm esc}(\leq1)$ of the galaxies is a power-law in redshift $z$ (also see Section \ref{sec:sims_mass_thresh}):
\begin{equation}\label{eq:f_esc}
    f_{\rm esc} = f_{\rm esc, 0}\bigg(\frac{1+z}{6}\bigg)^{\alpha_{\rm esc}},
\end{equation}
where $f_{\rm esc, 0}$ is the escape fraction normalisation and $\alpha_{\rm esc} = 0.20$ is the power-law index.

In this work, we adopt the same fiducial simulation as \cite{Balu_box}, L210\_AUG (hereon labelled Fiducial). This simulation has been calibrated with respect to the UV luminosity functions and the colour-magnitude relation within a rigorous Bayesian framework \citep{MERAXES-SNR-feedback}, as well as the stellar mass functions \citep{Balu_box}, at $z\sim 4 - 10$. Reionization parameters were tuned such that the reionization history is consistent with existing measurements of the IGM neutral fraction and the CMB optical depth \citep[see][in particular Figure 3 and Table 2]{Balu_box}.

\subsection{IGM Ionization State}\label{sec:excursion_set}

\MERAXES\:computes the IGM ionization, following the semi-numerical code \fastcm \citep{21cm-FAST}, via an excursion-set formalism \citep{Furlanetto_2004}. First, we grid the simulation volume and assign galaxies to the voxels based on their positions. We subdivide our simulation volume into $1024^3$ cells, corresponding to a cell size of $\sim 0.2~h^{-1}$ Mpc. In spheres of decreasing radii, we compare the number of ionizing photons from the stellar baryons and the total baryons in the IGM. After accounting for recombinations that can happen in the densest parts of the IGM, we flag a cell as ionized when the number of ionizing photons is higher than that of the neutral baryons. 
\begin{equation}\label{eq:excursion}
    N_{\rm b*} (\boldsymbol{x}, z | R) N_\gamma f_{\rm esc} \geq  N_{\rm atom} (\boldsymbol{x}, z | R) (1 + \bar{n}_{\rm rec}) (1 - \bar{x}_e),
\end{equation}
where $N_{\rm b*} (\boldsymbol{x}, z | R)$ is the number of stellar baryons in a sphere of radius $R$ centred at $\boldsymbol{x}$ and redshift $z$, $N_\gamma=4000$ is the number of UV ionizing photons per baryon \citep{Barkana_2007},  $N_{\rm atom}$ is the number of neutral \ion{H}{i} in the same volume, $\bar{n}_{\rm rec}$ is the average number of recombinations in the IGM \citep{Sobacchi2014}, and $\bar{x}_e$ is the mean electron fraction accounting for the secondary ionizations caused by X-ray photons. Motivated by the mean-free path of a typical UV photon in the IGM \citep{Songaila_2010}, we set the maximum of $R = 50$ Mpc and decrease it successively down to the size of a cell.

\subsection{21cm Signal}

The differential brightness temperature of the 21cm emission from a cloud of \ion{H}{i} gas illuminated by CMB radiation of temperature $T_{\rm{CMB}}$ is given by:
\begin{equation}
\label{eq:Tb-contrast}
\begin{split}
\delta T_{b} (\nu)  & = \dfrac{T_{\rm S} - T_{\rm{CMB}}}{1 + z} (1 - e^{-\tau_{\nu_0}})\\
&\approx  27 x_{\textsc{\ion{H}{i}}}(1 + \delta_{\rm nl} ) \left( \dfrac{H}{dv_{\rm r}/dr + H} \right) \left( 1 - \frac{T_{\rm{CMB}}}{T_{\rm S}} \right)\\
&\mathrm{\hspace{0.3cm}}\times \left( \frac{1+z}{10} \frac{0.15}{\Omega_{\rm M} h^{2}}\right)^{1/2} \left(\dfrac{\Omega_{\rm b} h^{2}}{0.023}\right)
\,\mathrm{mK},
\end{split}
\end{equation}
where $T_{\rm S}$ is the IGM spin temperature which determines the energy level populations of the \ion{H}{i} hyperfine states, $\tau_{\nu_0}$ is the optical depth, $\delta_{\rm nl} \equiv \rho/\Bar{\rho} - 1$ is defined as the evolved Eulerian density contrast ($\rho$ is the density), $H(z)$ is the Hubble parameter, $dv_r/dr$ is the line-of-sight co-moving velocity gradient, and $x_{\ion{H}{i}}$ is the neutral fraction. \MERAXES\:sources the density and the velocity fields from the N-body simulations and creates self-consistent $T_{\rm S}$ and $x_\ion{H}{i}$ fields.

\subsubsection{Spin Temperature}
As can be seen from the Equation (\ref{eq:Tb-contrast}), the spin temperature $T_{\rm S}$ of the IGM plays a major role in the 21cm signal. The level populations of the \ion{H}{i} hyperfine states depend on a number of physical processes in the IGM, including the amount and the energy of the UV and X-ray photons. $T_{\rm S}$ is related to the UV and X-ray emission via:
\begin{equation}\label{eq:spin_temperature}
    T_{\rm S}^{-1} = \dfrac{T_{\rm CMB}^{-1} + x_{\alpha} T_\alpha^{-1} + x_{c} T_{\rm K}^{-1}}{1 + x_{\alpha} + x_{\rm c}},
\end{equation}
where $x_\alpha$ and $x_{\rm c}$ are the Wouthuysen-Field coupling  \citep{Wouthuysen1952,Field1958} and the collisional coupling coefficients respectively. $x_{\rm c}$ is computed by taking into account the collisions of \ion{H}{i} atoms amongst themselves as well as with free electrons and protons in the IGM \citep{Zygelman,Furlanetto-Furlanetto}. $x_\alpha$ depends on the local \ion{Ly}{$\alpha$} background flux and closely follows the implementation in \citet{21cm-FAST}. $T_\alpha$ is the `colour' temperature, $T_{\rm K}$ is the kinetic temperature of the IGM, and we assume $T_\alpha = T_{\rm K}$ \citep{Field1959}.

The spin temperature field is therefore very sensitive to the $T_{\rm K}$, which is impacted by X-ray heating. The evolution of the $T_{\rm K}$ depends on the angle-averaged X-ray intensity  $J(\boldsymbol{x}, E, z)$ which is computed as a function of the position $\boldsymbol{x}$, X-ray photon energy $E$, and redshift $z$:
\begin{equation}
    J(\boldsymbol{x}, E, z) = \dfrac{(1+z)^3}{4\pi} \int_z^\infty dz' \frac{cdt}{dz^{'}} \epsilon_{\rm X} e^{-\tau},
\end{equation}
where we have integrated the comoving X-ray emissivity $\epsilon_{\rm X}$ back along the light cone, and $e^{-\tau}$ accounts for the probability that an X-ray photon emitted at redshift $z^\prime$ survives till $z$. We compute $\epsilon_{\rm X}$ as a function of the position $\boldsymbol{x}$, X-ray photon energy $E_e = E(1+z^\prime)/(1+z)$ at the emitted redshift $z^\prime$:

\begin{equation}
\epsilon_X(\boldsymbol{x}, E_e, z^\prime) = \dfrac{L_X}{\rm SFR} \times \rm{SFRD}(\boldsymbol{x}, E_e,  z'),
\end{equation}
where $L_X/{\rm SFR}$ is the galaxies' specific X-ray luminosity per unit star formation rate (SFR), and SFRD is the star formation rate density. We assume a power-law in X-ray photon energy $E$, $L_X/{\rm SFR} \propto E^{-\alpha_X}$, with $\alpha_X = 1$ which is consistent with observations of high mass X-ray binaries in the local Universe \citep{Mineo2012,Fragos_2013,Pacucci_2014}, and is normalised: 
\begin{equation}\label{eq:LXrayGal}
L_{X <2~\rm{keV}}/{\rm SFR} = \int_{E_0}^{2~\rm{keV}} dE_{e}~L_X/{SFR},
\end{equation}
where $L_{X <2~\rm{keV}}/{\rm SFR}$ is the soft-band X-ray luminosity per SFR in units of (erg s$^{-1}$ M$_\odot^{-1}$ yr), and $E_0$ fixes the minimum energy for an X-ray photon so that it is not absorbed within the galaxy. 

\subsection{Simulations}\label{sec:simulations}

\begin{table*}
\centering
\caption[]
{\small Astrophysical parameter summary for the seven \MERAXES\: simulations. See text for details.}
\begin{tabular}{@{}lccccl@{}}
\toprule
 Name & Minimum Halo Mass ($M_{\rm thresh}$) & $f_{\rm esc, 0}$ & X-ray Luminosity ($L_{X<2\,\rm{keV}}/\rm{SFR}$) & X-ray threshold ($E_0$) & Comments \\ 
 & $[M_{\sun}]$ & & $[\rm{erg\,s^{-1}\,M^{-1}_{\sun} \, yr}]$ & $[\rm{keV}]$ & \\ \midrule
 Fiducial & $3 \times 10^7$  & 0.14 & $3.16\times10^{40} $ & $0.5$ & Fiducial simulation \\
 \lowMh & $10^9$  & 0.25 &$3.16\times10^{40} $ & $0.5$ & Intermediate halo mass threshold simulation\\
 high\_$M_h$ & $10^{10}$ & 0.45 & $3.16\times10^{40} $ & $0.5$ & High halo mass threshold simulation \\
 low\_$L_X$ & $10^8$ & 0.14 & $3.16\times10^{38}$ & $0.5$ & Low X-ray luminosity simulation\\
 high\_$L_X$ & $10^8$ & 0.14 & $3.16\times10^{42}$ & $0.5$ & High X-ray luminosity simulation\\
 low\_$E_0$ & $10^8$  & 0.14 & $3.16\times10^{40} $ & $0.2$ & Low X-ray energy threshold simulation\\
 high\_$E_0$ & $10^8$  & 0.14 & $3.16\times10^{40} $ & $1$ & High X-ray energy threshold simulation\\ \bottomrule
\end{tabular}
\label{table:sim-params}
\end{table*}

To aid our physical interpretation of the features present in the SS we run a further set of six simulations with \MERAXES, in addition to our fiducial simulation.

In particular, we are interested in the physical processes which impact the morphology of the 21cm signal. To explore the impact of the ionization morphology we vary the minimum mass for halos hosting star formation. Setting the UV escape fraction to zero in galaxies below a given mass threshold alters the size and distribution of the ionized regions (i.e. produces larger, more isolated bubbles for an increasing mass threshold). With regard to the heating morphology, we vary the X-ray luminosity and the minimum energy threshold for X-rays escaping their host environment. Increasing the X-ray energy threshold decreases the prevalence of bubbles of heated IGM gas transitioning toward an effective uniform background of IGM heating. Table \ref{table:sim-params} summarises the simulations used in this work along with the values of the parameters that are varied.

\subsubsection{Halo Mass Threshold}\label{sec:sims_mass_thresh}

To explore the impact of the minimum halo mass on the EoR morphology, we modify the $f_{\rm esc}$ (also see Equation \ref{eq:f_esc} and Section \ref{sec:excursion_set}) prescription as follows:

\begin{equation} \label{eq:eta}
    f_{\rm esc} = \begin{cases}
    f_{\rm esc, 0}\bigg(\frac{1+z}{6}\bigg)^{\alpha_{\rm esc}}, & M_{\rm halo} \geq M_{\rm thresh} \\ \\ 
    0, & M_{\rm halo} < M_{\rm thresh}  .\\
  \end{cases}
\end{equation}

We run simulations with mass thresholds $M_{\rm thresh} = 10^9$ and $10^{10} \,M_\odot$, and label them \lowMh and \highMh respectively (we point out that our Fiducial simulation contains all haloes down to $3\times10^7\,M_\odot$). By considering an increasing halo mass threshold, we effectively decrease the total number of galaxies capable of contributing to reionization. To compensate for the loss of ionizing sources we increase the UV escape fraction of those remaining star-forming galaxies to ensure a reionization history consistent with our observational constraints. We therefore, increase the UV escape fraction normalisation $f_{\rm esc, 0}$ (see Equation \ref{eq:f_esc}) to $[0.25,0.45]$ respectively (see Fig \ref{fig:sim-neutral-frac}). These simulations can thus probe the impact of ionization morphology in the PS and SS. An increasing mass threshold should result in larger ionized regions (changing the physical location of features in the PS/SS). We emphasise that we still populate and evolve the galaxies in the haloes below the mass threshold, and these galaxies can start contributing to the UV ionization budget when their host halo mass passes the threshold. We point out that we do not suppress emission of X-ray photons by these galaxies. In this manner, we fix the X-ray background across these simulations (Fiducial, \lowMh and \highMh) to be the same. This was a deliberate design choice to isolate the impact of the EoR morphology on the 21-cm statistics.

\subsubsection{X-Ray Luminosity}

We also consider two simulations with a lower and higher X-ray luminosity $L_{X<2\,\rm{keV}}/\rm{SFR} = [3.16 \times 10^{38}, 3.16\times 10^{42}]\:\rm{ergs\,s^{-1}\, M_\odot^{-1} \,yr}$ as compared to the fiducial value of $3.16 \times 10^{40}\:\rm{ergs\,s^{-1}\, M_\odot^{-1} \,yr}$. These two simulations, labelled \lowLX and \highLX correspond to the L210\_AUG\_LOWX and L210\_AUG\_HIGHX simulations in  \citet{Balu_box}. These simulations cover a range of X-ray luminosities per SFR, one order of magnitude broader than what is observed in the local soft band X-ray luminosity \citep{Mineo2012,Fialkov2016} based on the range adopted in \citet{Greig2017}. 

For \lowLX, X-ray heating is inefficient and the IGM ionizes before it is heated (21cm signal always remains in absorption). This produces large temperature contrasts between the ionized and neutral regions resulting in much higher amplitudes for the 21-cm statistics.

\subsubsection{X-ray Energy Threshold}

We explore the impact of the X-ray photon energy threshold $E_0$ by producing two simulations with $E_0 = 0.2 \,\rm{keV}$ and $E_0 = 1 \,\rm{keV}$ compared to $E_0 = 0.5 \,\rm{keV}$ for the Fiducial simulation. Decreasing the energy threshold, coupled with our power-law X-ray spectral energy distribution, results in a higher fraction of softer X-ray photons. As softer photons have shorter mean free paths, more heat energy is deposited closer to the host galaxies resulting in more prevalent bubbles of heating around the first galaxies. Increasing the energy threshold removes this heating morphology as the X-ray photons now penetrate much deeper into the IGM before depositing their heat energy resulting in an effective uniform background of heating. In effect, varying this energy threshold will alter the amplitude of the 21-cm statistics during the heating epoch \citep[see e.g.][]{Pacucci_2014,Greig2017}.

\subsubsection{Fiducial no Spin Temperature}
\begin{figure*}
    \centering
    \includegraphics[width=\textwidth]{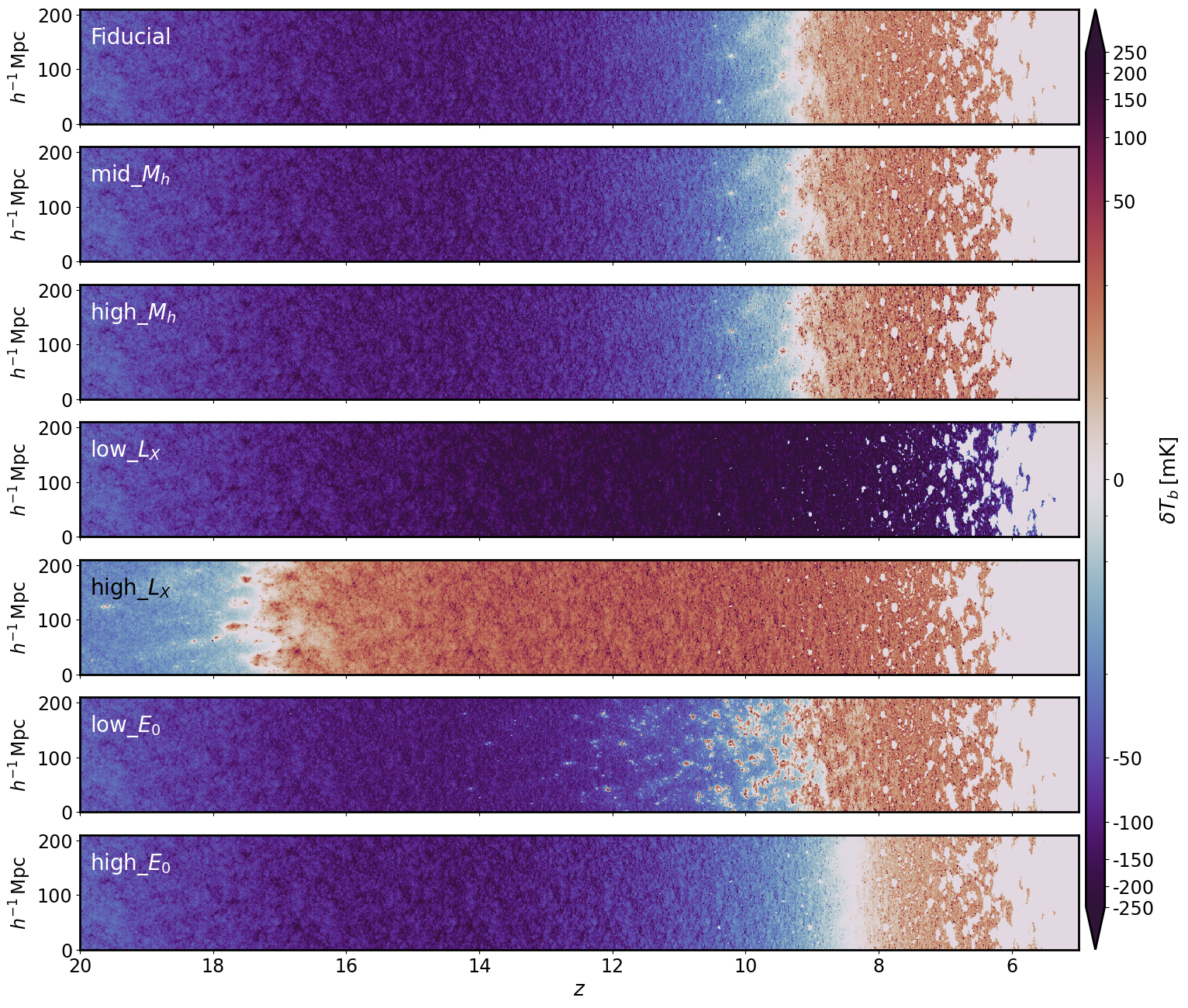}
    \caption{21cm differential brightness temperature light cone slices for each simulation volume as a function of redshift from $z=20-5$. The top slice is the Fiducial simulation (labelled), and the bottom slice is the \highEX simulation. The colour bar is a symmetric log-scale, where blue indicates absorption relative to the CMB, red indicates emission relative to the CMB, and gray indicates either ionization or zero signal. Each lightcone slice is fixed to the same temperature scale.}
    \label{fig:sim-lightcone-slices}
\end{figure*}
As a comparison we create an additional simulation of the Fiducial signal in the spin temperature saturation limit where $T_S \gg T_{\mathrm{CMB}}$ (from hereon \fidnoTS). This approximation effectively sets the temperature contrast independent of the spin temperature during reionization. In this limit the temperature field is proportional to the matter density $\delta_\rho$ and ionization field $x_{\mathrm{HI}}$ \citep{Cooray2005,Furlanetto2006b,Lidz_2007}. \citet{Shim_2017} and \citet{Maj2018} both explore the matter density and ionization bispectrum components to the total 21cm bispectrum in this limit. \citet{Maj2018} in particular finds that the negative sign of the bispectrum might be an important indicator of the ionization topology during reionization.

\section{Results}\label{sec:Stats-thermal-history}

In this section, we analyse the thermal cosmological history of the 21cm brightness temperature signal for the simulation sets. We plot 2D slices of the lightcone boxes as a function of redshift for each simulation. We also calculate the neutral fraction $(\Bar{x}_{\textrm{HI}})$ for each simulation coeval box as a function of redshift, and compare the results of each simulation. We then calculate the PS and SS for each simulation as a function of redshift for large $(k\sim0.1\,\mathrm{Mpc}^{-1})$ and small $(k\sim1\,\mathrm{Mpc}^{-1})$ spatial scales.

\subsection{21cm Lightcones}\label{sec:21cm-lightcones}

%
Fig \ref{fig:sim-lightcone-slices} shows a lightcone slice of each simulation as a function redshift. In descending order the panels show the Fiducial, \lowMh, \highMh, \lowLX, \highLX, \lowEX, and \highEX simulations. The colour bar is a log symmetric colour map, where blue indicates absorption, and red indicates emission relative to the CMB. Gray indicates zero temperature difference. During reionization $\delta T_b = 0$ is typically associated with regions that are ionized.

The same $N$-body dark matter particle \GENESIS simulations are used to generate each of the different \MERAXES\:simulations. Therefore each simulation has the same dark matter halo distribution. In Fig \ref{fig:sim-lightcone-slices} this is evident at high redshifts ($z\lesssim20$) and during reionization ($z\lesssim8$) in the location and approximate size of the first ionization regions. There are some obvious differences in the temperature contrast due to the different X-ray heating parameters. Of note, we see that the \lowLX is always in absorption, even during reionization, and the \highLX simulation is heavily preheated at high redshift. The \lowEX simulation has small regions of localised heated gas that appear in emission at $z>15$. In contrast, the \highEX simulation results in a more uniform heating of the IGM and thus the brightness temperature is relatively featureless. There is also a clear difference in the size of ionization regions between the Fiducial, \lowMh, and \highMh simulations, with the size increasing from Fiducial to \highMh at fixed redshift. We discuss these features in the context of the statistics in the following subsections.

\subsection{Neutral Fraction and Ionization}\label{sec:neutral_fraction}

\begin{figure}
    \centering
    \includegraphics[width=0.475\textwidth]{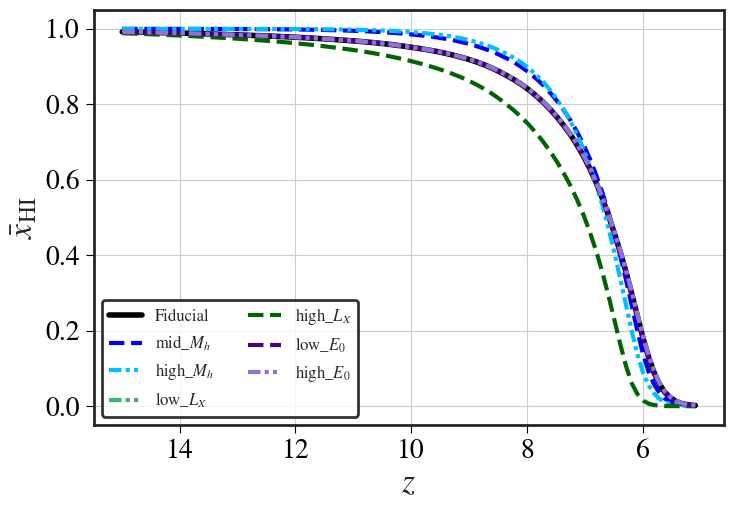}
    \caption{The ionization history of the average neutral hydrogen (neutral fraction $\Bar{x}_{\textrm{HI}}$) IGM calculated for each simulation (labelled).}
    \label{fig:sim-neutral-frac}
\end{figure}

In Fig \ref{fig:sim-neutral-frac} we show the average neutral fraction ($\Bar{x}_{\textrm{HI}}$) calculated for each simulation coeval box as a function of redshift. The Fiducial model was calibrated to match existing observational constraints (see \citet{Balu_box} for details). The halo mass threshold simulations have increased UV escape fractions as a function of halo mass to ensure similar reionization histories for easier comparison of the ionization morphology. The solid black line is the average neutral fraction for the Fiducial simulation. By $z=10$ the Fiducial simulation is already partially ionized at the $\sim5$ percent level. We note that the ionization history for the \lowEX (purple dashed line), the \highEX (light purple dash dotted line), and the \lowLX (light green dash dotted line) simulations are effectively identical to the Fiducial case. This is expected, since ionization is predominantly driven by UV photons not X-ray emission, additionally these simulations also have the same halo mass thresholds and escape fraction $(f_{\rm esc, 0})$ as the Fiducial simulation (see Table \ref{table:sim-params}). In contrast the \highLX (dark green dashed line) simulation undergoes reionization early relative to the Fiducial, due to the increase in the number of ionizations following secondary collisions of the X-ray photons. The \highLX simulation has a ionization fraction of $\sim10$ percent reionization by $z=10$. This is not unexpected since X-ray emission can be responsible for at most $10$ percent of ionization \citep{Mesinger2013}. 

The \lowMh (dark blue dashed line) and the \highMh (light blue dash dotted line) simulations begin reionization later than the Fiducial simulation, as it takes longer for haloes to gravitationally grow in excess of their respective mass thresholds to emit ionizing UV photons. Nevertheless, the \lowMh and the \highMh simulations neutral fraction profiles result in a similar ionization history to the Fiducial. This is due to the effective parameterisation of the escape fraction relative to the halo mass threshold. To compensate for the loss of ionizing sources as a function of increasing the halo mass threshold, the UV escape fraction was increased proportional to the halo mass threshold. The \lowMh and \highMh simulations have UV $f_{\rm esc, 0}$ values of $0.25$, and $0.45$, compared to the Fiducial with $0.14$. Therefore, more ionizing UV photons escape per unit mass from the same higher mass halos in \highMh and \lowMh, for the same amount of star formation, compared to the Fiducial simulation. As reionsation progresses, star formation increases, this results in a more rapid (sharper) reionization relative to the Fiducial simulation.

\subsection{21cm Statistics}\label{sec:21cm-statistics}

\subsubsection{Mean Brightness Temperature}

\begin{figure}
    \centering
    \includegraphics[width=0.475\textwidth]{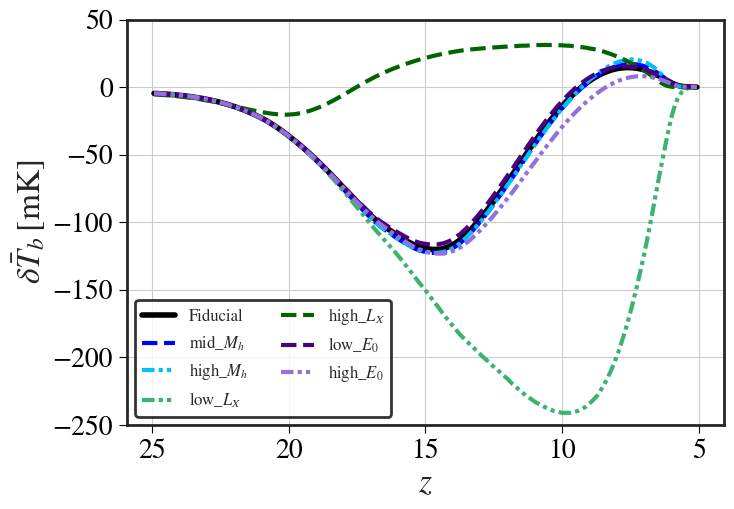}
    \caption{Mean brightness temperature for all simulations, calculated from the coeval boxes as a function of redshift.}
    \label{fig:all_sims_meanTb}
\end{figure}

Fig \ref{fig:all_sims_meanTb} shows the mean temperature for each simulation as a function of redshift. As previously mentioned, the Fiducial model, \lowLX and \highLX simulations are taken from \citet{Balu_box}. We see that the mean temperature for these three simulations agree with those shown in Figure 8 of \citet{Balu_box}, for more details on the \lowLX and \highLX simulations we refer the reader to this work. For all simulations we see a characteristic absorption feature which occurs when Ly-$\alpha$ emission couples the spin temperature to the gas temperature. As the gas expands adiabatically it cools relative to the CMB, increasing the relative absorption. For most simulations this absorption trough occurs at approximately $z\sim15$ (with the exception of the \highLX simulation). The timing of the trough depends on X-ray heating which eventually drives the IGM into emission (with the exception of the \lowLX simulation), occurring during reionization at $z\gtrsim10$. This culminates in a peak roughly at the midpoint of reionization in the redshift range of $z\sim6-8$.

\begin{figure*}
    \centering
    \includegraphics[width=\textwidth]{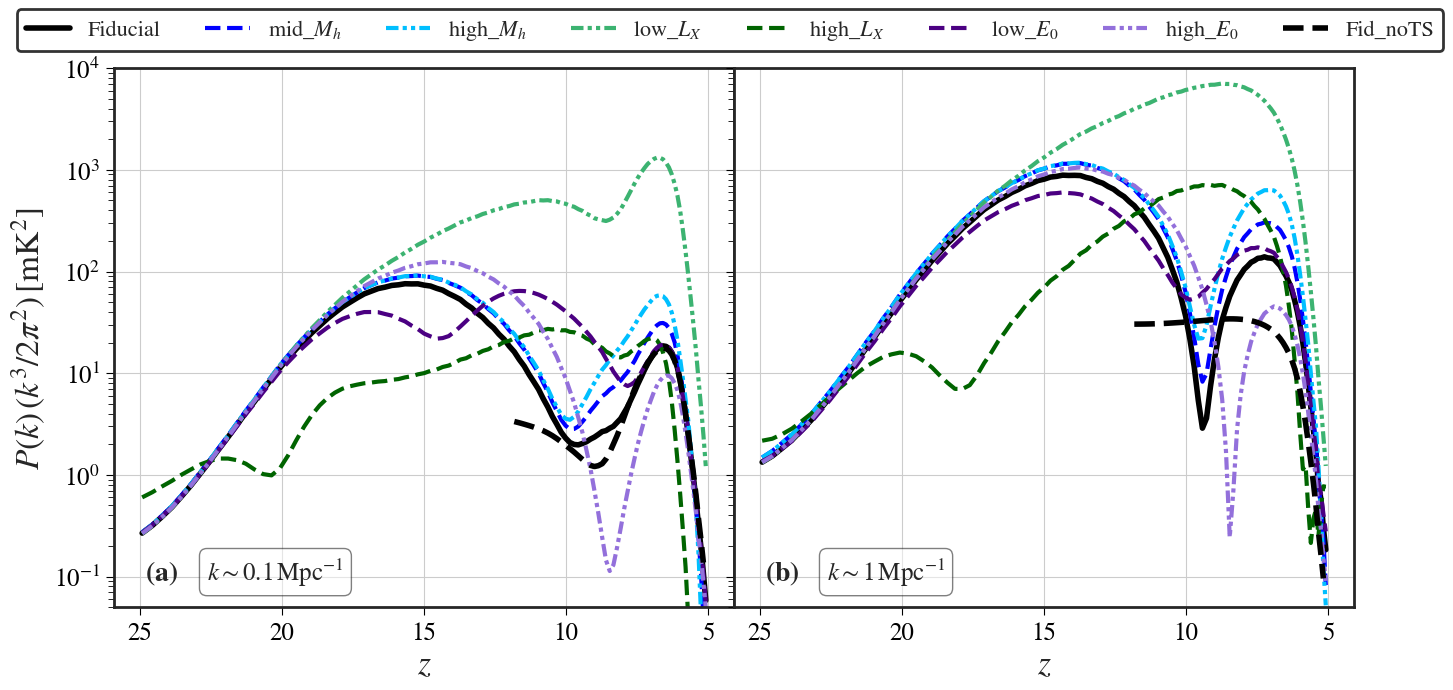}
    \caption{The PS as a function of redshift for all simulations at a fixed spatial scale. Panel (a) shows the redshift evolution at $k\sim0.1\,\rm{Mpc}^{-1}$, and panel (b) shows the redshift evolution at $k\sim1\,\rm{Mpc}^{-1}$.}
    \label{fig:all_sims_PS_large_2_small_k}
\end{figure*}

As mentioned in the previous section the halo mass simulations both start reionization later, but conclude earlier than the Fiducial. This delay results in a higher heating and reionization rates for the halo mass simulations. The higher ionization rate and heating being directly associated with the larger escape fraction. This directly affects the mean temperature and the amplitude of the PS (as seen in Fig \ref{fig:all_sims_PS_large_2_small_k}).

The X-ray energy threshold simulations follow a similar evolution to the Fiducial. The biggest difference is between the amplitude and onset of X-ray heating between $10 < z <15$. Here the \lowEX (purple dashed line) undergoes heating earlier than the Fiducial and the \highEX (light purple dash dotted line) simulations. This should be expected since there are relatively more lower energy soft X-rays available to heat the IGM. Additionally the timing of heating should be earlier for the \lowEX simulation because the mean free path of X-ray photons is proportional to their energy, meaning softer X-rays deposit their energy into the IGM before harder X-rays. 

\subsubsection{21cm Power Spectrum}

Fig \ref{fig:all_sims_PS_large_2_small_k} shows the PS calculated at spatial scales $k\sim0.1\,\mathrm{Mpc}^{-1}$ (panel (a)), and $k\sim1\,\mathrm{Mpc}^{-1}$ (panel (b)), as a function of redshift for each coeval box, and each simulation. The lines in Fig \ref{fig:all_sims_PS_large_2_small_k} correspond to the same simulations in Fig \ref{fig:all_sims_meanTb}. The features discussed in Fig \ref{fig:all_sims_meanTb} are broadly mirrored in Fig \ref{fig:all_sims_PS_large_2_small_k} (a) for most simulations. We find the Fiducial simulation in panel (a) of Fig \ref{fig:all_sims_PS_large_2_small_k} at $k\sim0.1\,\rm{Mpc}^{-1}$ has a peak at $z\sim16$ during Ly-$\alpha$ pumping \citep{Wouthuysen1952,Field1958}, and a peak at $z\sim6.5$ during the mid point of reionization. Most simulations do not have the characteristic three peak structure seen in \citep{Furlanetto2006,Mesinger2013}, with the exception of the \lowEX simulation and the \highLX simulation, the latter which has four peaks. The peak during the EoH for the Fiducial simulation has mostly merged with the peak during the EoR due to the delayed heating.

We also include the PS as a function of redshift for the \fidnoTS simulation (dashed black line). We only show the \fidnoTS simulation up to $z=12$, since the approximation is only valid during reionization. We find good agreement between the Fiducial simulation and the \fidnoTS simulation during the middle and late periods of reionization at $k\sim0.1\,\rm{Mpc}^{-1}$. At the larger scales the ionization topology is the dominant component in the PS. At smaller scales X-ray heating and other coupling effects are important for setting the spin temperature. These differences result in disagreement at $k\sim1\,\rm{Mpc}^{-1}$ since the spin temperature calculation is omitted in the \fidnoTS simulation.

The \lowMh and the \highMh simulations in panel (a) have broadly the same PS, differing at most in amplitude during the EoR, with the \highMh simulation peaking earlier than the \lowMh simulation. The \lowMh and \highMh simulations compared to the Fiducial have reionization topologies driven by larger (and more biased) sources. This results in an increase in the amplitude of the 21cm PS. The PS of both simulations are broadly the same as the Fiducial simulation, only significantly deviating at $z\sim10$ when reionization begins to become more significant for the Fiducial simulation. 

For the \lowLX and \highLX simulations, we find the same features at $k\sim0.1\,\rm{Mpc}^{-1}$ and $k\sim1\,\rm{Mpc}^{-1}$ in Fig~\ref{fig:all_sims_PS_large_2_small_k} as seen in \citet{Balu_box} for the L210\_AUG\_lowX and L210\_AUG\_highX simulations\footnote{The \lowLX and \highLX simulations are the L210\_AUG\_lowX and L210\_AUG\_highX from \citet{Balu_box}.}. The high amplitude of the \lowLX simulation is due to the higher temperature contrast that results from the colder IGM (lower X-ray luminosity). The lower amplitude of the \highLX simulation is due to the higher X-ray luminosity heating, this reduces the temperature fluctuations on all scales. The inefficient heating due to lower X-ray luminosity in the \lowLX simulation results in the merging of the EoH and EoR peaks. In the \highLX simulation we have four peaks, an early peak at $z\sim22$ coincident with the relatively weak absorption trough (Ly-$\alpha$ pumping), a peak at $z\sim18$ which corresponds to the EoH heating, with the IGM being in emission at this stage. There is a peak during the midpoint of reionization at $z\sim7$. In addition to the expected three peaks there is an additional peak at $z\sim10$. This is due to the first ionization sources, with the \highLX simulation having $\sim10$ percent ionization by $z=10$. 

For the \highEX case at $k\sim0.1\,\rm{Mpc}^{-1}$ the first peak occurs at $z\sim14$, which is roughly coincident with the absorption trough in the mean temperature brightness in Fig~\ref{fig:all_sims_meanTb} which reaches a minimum at $z\lesssim14$. The second less prominent peak occurs during the EoR ($z\sim6$). The lower amplitude and later occurrence of the peak during the EoR are a direct result of the higher X-ray energy threshold, which produce less structure in the IGM due to the more uniform heating of the longer mean-free path X-ray photons. The first patches of emission that are correlated on the largest scales appear later relative to the other simulations, we see similar behaviour in Figure 1 of \citet{Greig2017} (bottom row of Figure 1). This delayed and more uniform heating means the Ly-$\alpha$ pumping dominates the amplitude of the PS for a longer period. 

\begin{figure*}
    \centering
    \includegraphics[width=\textwidth]{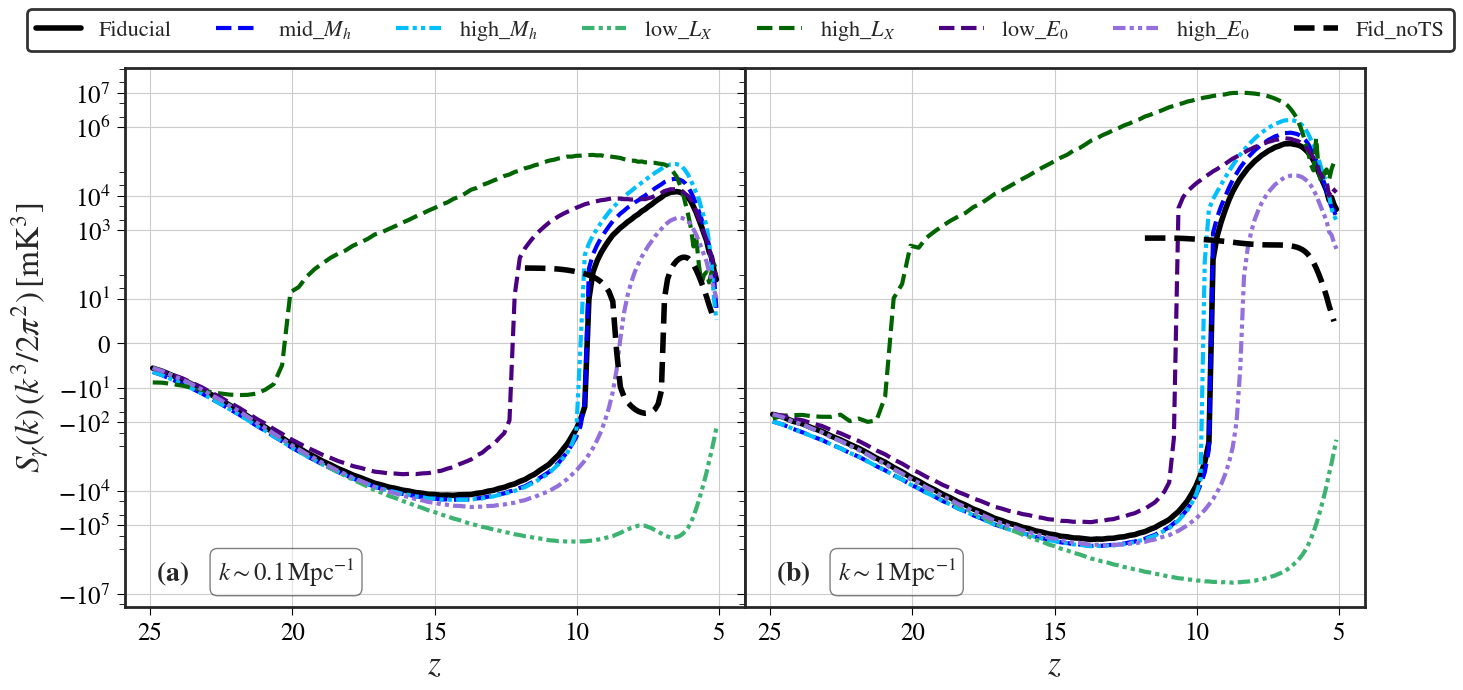}
    \caption{The SS as a function of redshift for all simulations at a fixed spatial scale. Panel (a) shows the redshift evolution at $k\sim0.1\,\textrm{Mpc}^{-1}$, panel (b) shows the redshift evolution at $k\sim1\,\textrm{Mpc}^{-1}$.}
    \label{fig:all_sims_SS_large_2_small_k}
\end{figure*}

For the \lowEX simulation at fixed scale $k\sim0.1\,\rm{Mpc}^{-1}$ we see the three peaked structure. The peak due to Ly-$\alpha$ occurs early at $z\sim17$, with the peak during the EoH occurring at $z\sim12$ which coincides with the pockets of emission seen in Fig \ref{fig:sim-lightcone-slices}. The amplitude of the EoH is greater for the \lowEX simulation due to the lower X-ray energy threshold. These lower energy X-rays efficiently heat the local medium around the first luminous sources, producing higher temperature contrasts in the IGM due to the more inhomogeneous heating. This produce more inhomogeneous structures, increasing the overall power. These correlate on the largest scales resulting in a strong peak. The high temperature contrast regions are the first to ionize, thus during the EoR, the relative amplitude of the EoR peak returns to that of the Fiducial model.

\subsubsection{21cm Skew Spectrum}\label{sec:skew-spectrum-redshift}
Fig \ref{fig:all_sims_SS_large_2_small_k} shows the SS calculated at $k\sim0.1\,\rm{Mpc}^{-1}$ (panel (a)), and $k\sim1.0\,\rm{Mpc}^{-1}$ (panel (b)), as a function of redshift for each coeval box, and each simulation. The lines correspond to the same simulations in Fig \ref{fig:all_sims_meanTb} and \ref{fig:all_sims_PS_large_2_small_k}. 

The SS for each simulation, broadly mirrors the mean brightness temperature for both small (panel (b)) and large (panel (a)) spatial scales in Fig \ref{fig:all_sims_SS_large_2_small_k}. The transition to a positive SS occurs rapidly within one coeval redshift box (a cosmic blink). This happens in all simulations except \lowLX which is always in absorption\footnote{We do not include the \fidnoTS in this paragraph, since it is only relevant during the EoR.}. The transition to positive SS for almost all simulations occurs earlier ($\Delta z < 1$) and more rapidly than the mean temperature brightness. Looking at Fig \ref{fig:sim-lightcone-slices} there is a clear explanation. During the transition period between the EoH and the EoR ($z\sim10$), the star formation rate increases. The X-ray luminosity is proportional to the star formation rate. This results in the appearance of heated islands that are in emission relative to the rest of the IGM which is undergoing more uniform and less efficient heating. These heated islands skew the temperature distribution towards positive values, resulting in a positive SS. This eventually tapers as the IGM saturates and the heated regions overlap, completing the transition from absorption to emission.

This transition happens significantly earlier for the \lowEX simulation ($\Delta z\sim2.5$). Due to the lower X-ray energy threshold, the \lowEX simulation undergoes earlier and more intense local heating. The amplitude of the effect appears to be scale dependent, with the transition in panel (b) at smaller scales occurring much closer in redshift to the transition in the mean temperature brightness. Conversely, for the \highEX simulation we see a $\Delta z\sim1$ lag in this transition at large and small scales. This is due to the delayed and more uniform heating from the high X-ray energy threshold.

During reionization, the ionization morphology appears to dominate the signal in the SS, with a peak associated with the midpoint of reionization, with the exception of the \lowLX simulation which has a trough (effectively mirrored about the x-axis). As the neutral fraction drops below $50$ percent most of the medium is ionized, and this tapers the SS, much like the PS, driving the signal to zero as reionization progresses. Like the PS, the SS amplitude is largest for the \highMh simulation at the largest scale due to the fact that reionization is driven by larger, more biased galaxies. However, the \highLX simulation has the highest overall amplitude at all scales in Fig \ref{fig:all_sims_SS_large_2_small_k}. The efficient X-ray heating in the \highLX simulation drives the non-Gaussianities, and results in an earlier reionization, and therefore a peak at ($z\sim7$) compared to the other simulations. 

\begin{figure*}
    \centering
    \includegraphics[width=\textwidth]{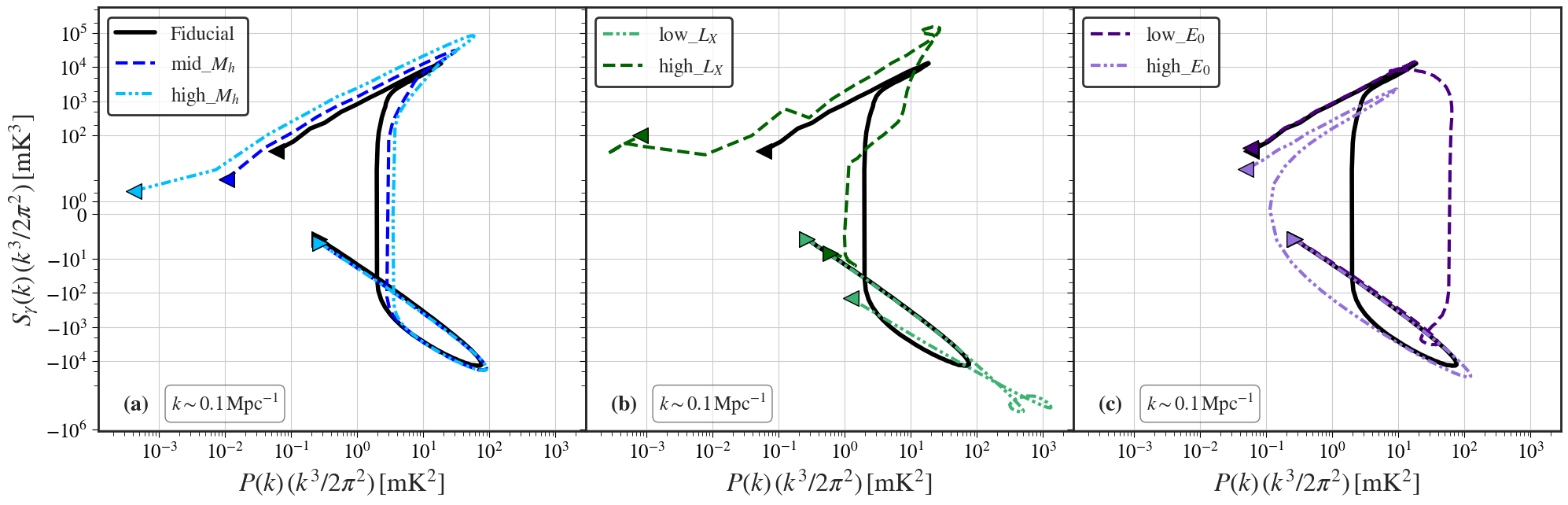}
    \caption{The dimensionless SS plotted against the dimensionless PS as a function of redshift for each simulation at a fixed spatial scale $k\sim0.1\,\rm{Mpc}^{-1}$. Each curve indicates the co-evolution of the PS and SS amplitude as a function of redshift, with the right pointed triangle, indicating the start point at $z=20$, and the left pointing triangle indicating the endpoint at $z=5$.}
    \label{fig:PS_SS_loci}
\end{figure*}

%
%
For contrast we calculate the SS for the \fidnoTS simulation for all redshifts up to $z=12$ (black dashed line). At large scales ($k\sim0.1\,\mathrm{Mpc}^{-1}$) we find a dip in the SS from positive to negative. This feature agrees with similar work performed by \citet{Maj2018} who look at different bispectrum of the expected 21cm signal in the saturation limit ($T_S \gg T_{\mathrm{CMB}}$). This transition occurs because the first ionization regions appear, which have zero signal and skew the distribution in the negative direction (growing peak at zero in an otherwise positive distribution). This result clearly contrasts with the other simulations, of particular note is the difference in amplitude with \fidnoTS having $2-3$ orders of magnitude lower SS amplitude in comparison. This illustrates the importance of the spin temperature in the higher order statistics.

%
%
MP23 investigate the SS during the EoH and the EoR with \fastcm. In Figure 3 and 4 of MP23, they display the SS as a function of redshift at different fixed scales (small and large) for several different simulations. In general they find two positive peaks in the SS across all spatial scales for all simulations. MP23 associate the first peak with X-ray heating which couples the spin temperature to the matter density. However, the transition from negative SS to positive occurs at $z=14-12$ for the simulations presented in MP23. We find this transition is likely a result of the appearance of the first ionization regions, since at this redshift range the simulations are at the minimum temperature (maximal absorption relative to the CMB). The second transition from positive to negative occurs at $z\sim10$ when the neutral IGM goes from absorption to emission, effectively flipping the temperature distribution and its asymmetry about the y-axis. The final transition happens when the IGM is more than $50$ percent ionized, in this case the remaining neutral medium (which is in emission) skews the temperature distribution towards positive values.

The simulations in MP23 have effectively the same X-ray parameters as the Fiducial simulation in our work. However, we do not find a negative SS at the start of reionization, instead we see a single peak and trough for the Fiducial, \lowMh, \highMh and \highEX simulations. Furthermore, we note that the absolute amplitude of our SS for all simulations is $10-10^2$ times large than those in MP23. At the largest scale $k=0.1\,\mathrm{Mpc}^{-1}$ for the \lowEX, the \highLX and the \lowLX simulations we do find some weaker peak and trough features towards the earlier stages of reionization $(z\sim7.5)$, when the \fidnoTS simulation SS is negative. The absolute change in amplitude as a result of these features are on the order of $10^3-10^5\,\mathrm{mK}^3$. This is comparable to the absolute change in MP23 that results in the a sign transition. The lack of a sign transition for the \MERAXES\:simulations, in contrast to MP23 and the \fidnoTS, is a clear indicator, that the spin temperature is inherently far more skewed than in the latter cases. We further discuss the differences, and the sources of this larger skewness in \MERAXES\:in Section \ref{sec:sign-flip-disc}.
%
%

\subsubsection{Skew Spectrum and Power Spectrum Co-evolution}

Fig \ref{fig:PS_SS_loci} shows the dimensionless SS and PS of each simulation as a function of redshift. Each simulation set is separated into individual panels, with the \lowMh, and \highMh in panel (a), the \lowLX and \highLX in panel (b), and the \lowEX, and \highEX simulations in panel (c). Each curve is taken at $k\sim0.1\,\rm{Mpc}^{-1}$, the colour coded triangles show the direction the curve travels as a function of redshift. The start point is indicated by the right facing triangle (relative to the peak), and the endpoint being the left facing triangle.

The curves in Fig \ref{fig:PS_SS_loci} demonstrate how the SS and PS co-evolve as a function of redshift. For the Fiducial simulation we see the PS increases as the SS becomes more negative, this occurs during absorption at early redshifts. Then as the IGM heats due to X-ray emission, the PS amplitude decreases, and the SS amplitude increases towards zero, until eventually becoming positive at a fixed PS amplitude. The PS and SS then both increase as reionization and heating occur in tandem, eventually culminating in a downward trend towards zero after the midpoint of reionization. This evolution is broadly mirrored by the other simulations with the exception of the \lowLX and \highLX simulations. 

For the halo mass simulations, the co-evolution of the SS and PS as a function of redshift is practically identical compared to the Fiducial simulation. The deviations occur due to the differences between the PS and SS amplitudes. During the EoH since all simulations have the same X-ray background, the main difference is due to heating from UV emission. For the Fiducial simulation this can occur at lower halo masses. The heating leads to a reduction in the PS amplitude for the Fiducial simulation when the IGM is in absorption; this effect is however minimal compared to X-ray heating \citep{Furlanetto2006b}. Once reionization starts, the \lowMh and \highMh simulations can emit more UV photons per unit mass per unit star formation, this results in relatively more UV heating. When the IGM is in emission, this leads to higher SS and PS amplitudes for the \lowMh and \highMh simulations compared to the Fiducial.

For the \lowLX simulation the SS is always negative, there is also a fairly flat SS from $z=10$ to $z=6$, with a small peak. These corresponds to the turning points in the curve, similar to the Fiducial simulation. The \highLX simulation demonstrates deviation from the Fiducial evolution, with the turning point from negative to positive SS amplitude happening at low PS amplitude. We see more turning points with increasing PS amplitude for the \highLX simulation which are correlated with the two peaks seen between $10<z<20$ in Fig \ref{fig:all_sims_PS_large_2_small_k} panel (b). These finally culminate in a turning point at high SS and PS amplitude during the peak of reionization, transitioning to the zero amplitude for both.

\begin{figure*}
    \centering
    \includegraphics[width=\textwidth]{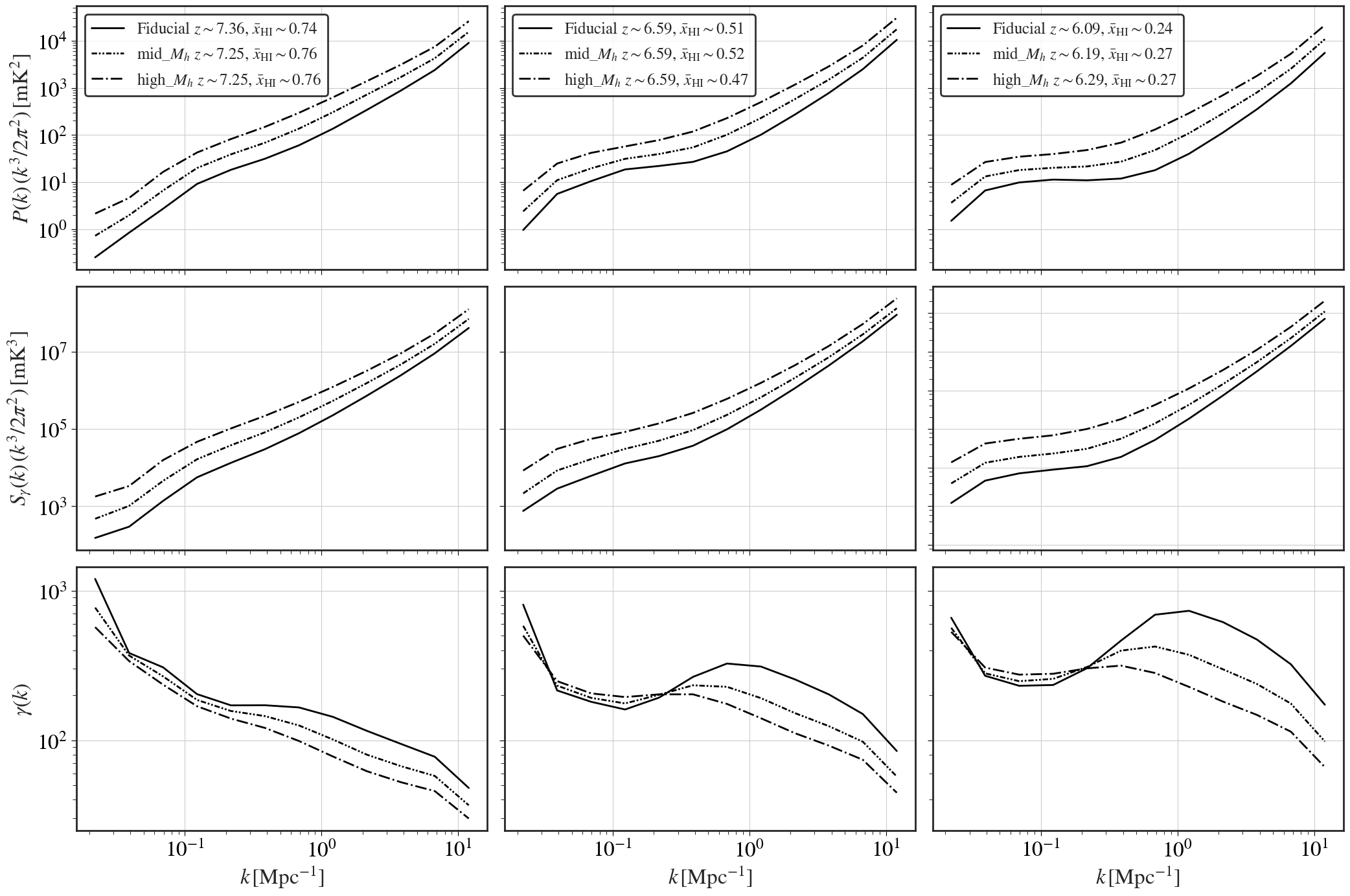}
    \caption{PS (row one), SS (row two) and normalised SS (row three), of the Fiducial (solid black line), \lowMh (double dotted dash line), and the \highMh (dash dotted line). Each figure from left to right is the $\Bar{x}_{\textrm{HI}}\sim 0.75,\:0.5,$ and $0.25$ for each simulation.}
    \label{fig:PS_SS_Gamma_Mh}
\end{figure*}

For the \lowEX and \highEX simulations, the evolution is notably different. For the \lowEX simulation, the structure mirrors the Fiducial simulation, with the sharp transition in amplitude for the SS happening earlier and at higher PS amplitudes due to the increased heating morphology. The more uniform heating of the IGM in the case of the \highEX simulation leads to less variation, and thus to less rapid growth of the SS and PS amplitude with respect to the other simulations. Notably, the transition from negative to positive SS amplitude occurs over a longer period for \highEX. The PS amplitude in this case varies as the SS amplitude transitions from negative to positive; this is a key difference between the \highEX on other simulations.

\section{Power Spectrum and Skew Spectrum During reionization}\label{sec:PS-SS-EoR}

In this section we calculate the spherically averaged PS and the spherically averaged SS for each simulation during the epoch of reionization. We investigate what information we are sensitive to during reionization, and what measuring the PS and the SS together can reveal about the physics of reionization. We investigate this by measuring the normalised SS, discussed in the following section \ref{sec:normalise-ss}.

\subsection{Normalised Skew Spectrum}\label{sec:normalise-ss}

The SS like the bispectrum is a measure of the non-Gaussianity through the central third order moment statistics of the temperature field. \citet{Wat2018} found significant fluctuations in the expected bispectrum around the zero point. These large fluctuations are due to the PS amplitude present in the statistic. Inspired by \citet{Eggemeier2016}, \citet{Wat2018} normalises the bispectrum to a unitless `normalised' bispectrum, which is normalised by the PS and the $k$-modes. This normalisation is akin to measuring skewness, which is the central third order moment normalised by the cube of the standard deviation (the variance to the power of $3/2$). In this work we perform a similar normalisation to remove the Gaussian component of the amplitude in the SS. We normalise the SS by the PS taken to the power of $3/2$, providing a unitless quantity $\gamma(k)$, referred to as the normalised SS:
\begin{equation}\label{eq:[psuedo-skewness]}
    \gamma(k) = \frac{\Delta^2_{T^2T} }{\left(\Delta^2_{T}\right)^{3/2}}.
\end{equation}

\begin{figure*}
    \centering
    \includegraphics[width=\textwidth]{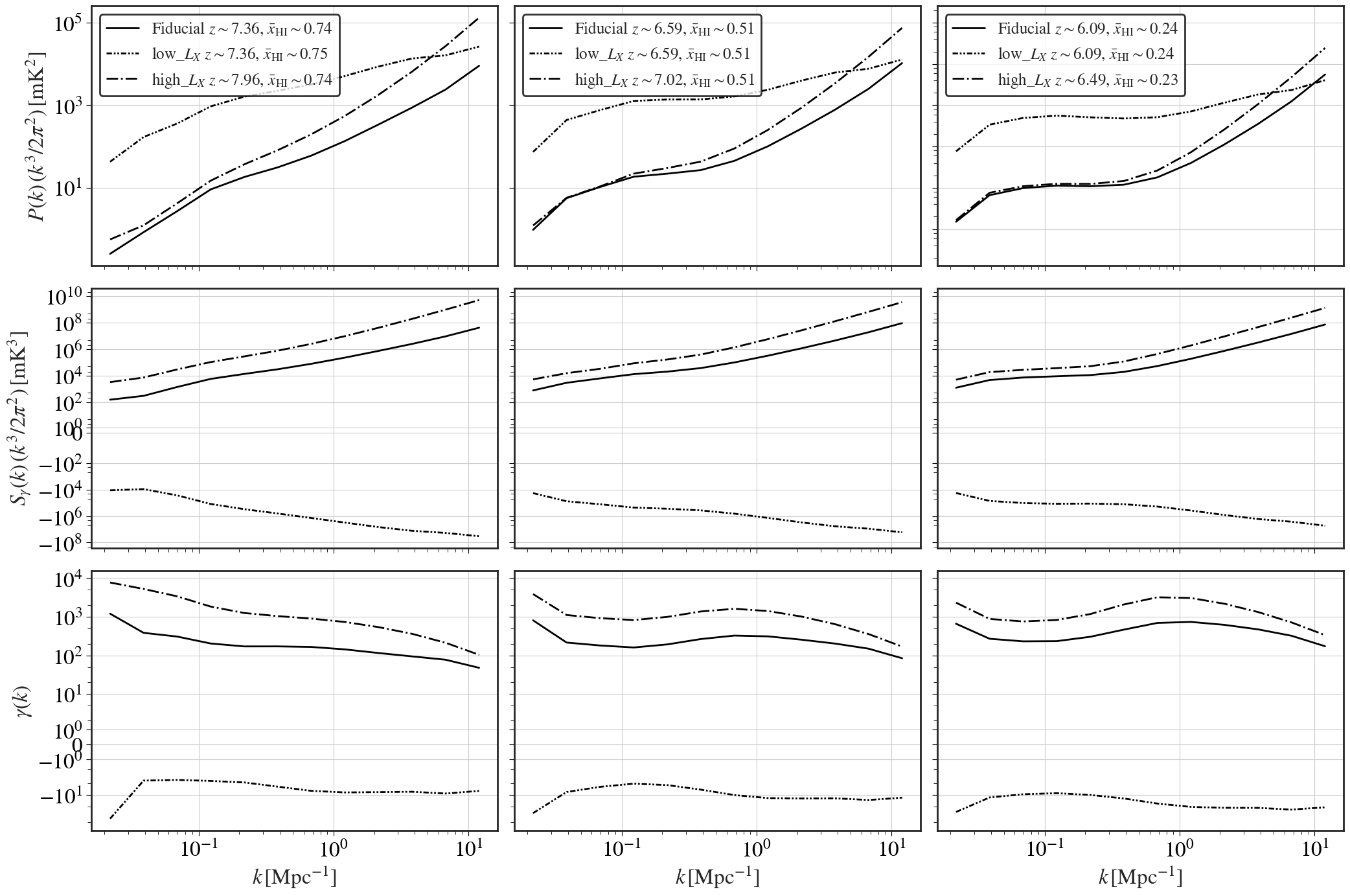}
    \caption{PS (row one), SS (row two) and normalised SS (row three), of the Fiducial (solid black line), \lowLX (double dotted dash line), and the \highLX (dash dotted line). Each figure from left to right is the $\Bar{x}_{\textrm{HI}}\sim 0.75,\:0.5,$ and $0.25$ for each simulation.}
    \label{fig:PS_SS_Gamma_LX}
\end{figure*}

In Equation \ref{eq:[psuedo-skewness]} $\Delta^2_{T}$ and $\Delta^2_{T^2T}$ are the dimensionless spherically averaged PS and SS, which have units of $\textrm{mK}^2$ and $\textrm{mK}^3$ respectively. The subscripts $T$ and $T^2T$ indicate the dimensionless PS and SS respectively. Deviations from a flat distribution as a function of spatial scale will be indicative of when the PS of the temperature fluctuations is more or less significant relative to the SS. 

\begin{figure*}
    \centering
    \includegraphics[width=\textwidth]{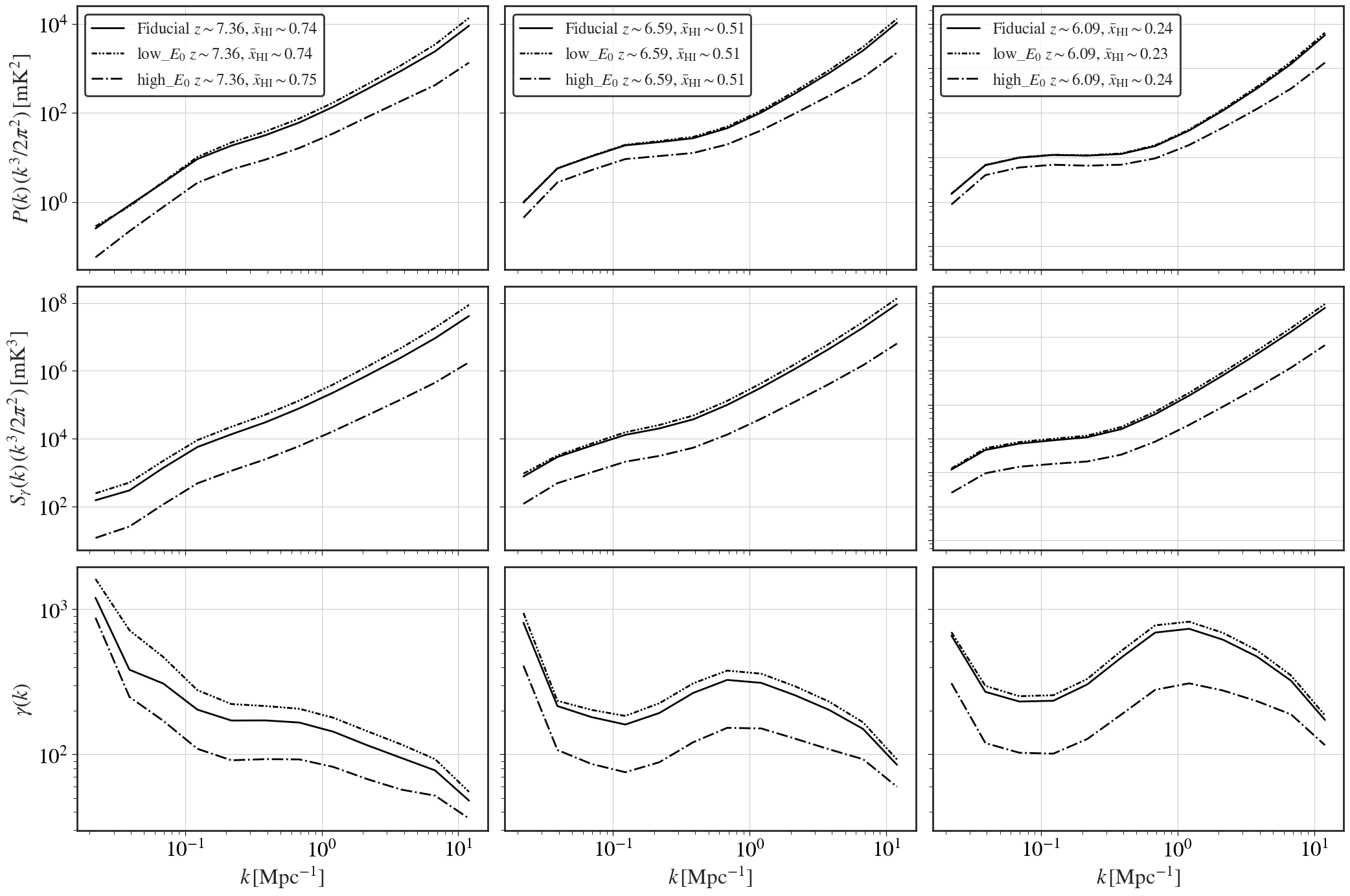}
    \caption{PS (row one), SS (row two) and normalised SS (row three), of the Fiducial (solid black line), \lowEX (double dotted dash line), and the \highEX (dash dotted line). Each figure from left to right is the $\Bar{x}_{\textrm{HI}}\sim 0.75,\:0.5,$ and $0.25$ for each simulation.}
    \label{fig:PS_SS_Gamma_E0}
\end{figure*}

\subsection{Power Spectrum, Skew Spectrum, and the Normalised Skew Spectrum}\label{sec:PS_SS_gamma}


In this section we focus our investigation on what the PS and SS look like for each simulation at different stages of reionization, specifically at neutral fractions of $x_{\rm{HI}} \sim 0.75,\,0.5$, and $0.25$. This allows for a direct comparison of the state of the IGM for each simulation set as compared to the Fiducial simulation. In addition to the PS and SS at these states of the IGM, we also calculate the normalised SS $(\gamma(k))$ from Equation \ref{eq:[psuedo-skewness]} for each simulation. $\gamma(k)$ allows for the isolation of the non-Gaussianity of the signal by normalising out the PS amplitude. In Figs \ref{fig:PS_SS_Gamma_Mh}, \ref{fig:PS_SS_Gamma_LX} and \ref{fig:PS_SS_Gamma_E0} we show the spherically averaged PS (first row), the spherically averaged SS (second row), and $\gamma(k)$ (third row) as a function of spatial scale, with the different neutral fractions of $\sim0.75$, $\sim0.5$ and $\sim0.25$ representing the first, second and third columns of each figure. 

In Fig \ref{fig:PS_SS_Gamma_Mh} we look at the PS, SS and the normalised SS of the halo mass simulation set. In Fig \ref{fig:PS_SS_Gamma_Mh} the Fiducial simulation is the solid black line, the \lowMh the double dot dashed line, and the \highMh is the dashed dot line. For both the PS and the SS we see a flattening of the spectra ($k<1 \,\rm{Mpc}^{-1}$) as reionization progresses. This is typically related to the ionization morphology \citep{Zaldarriaga_2004}. Notably, the PS and SS are very similar as a function of $k$, this is due to the Gaussian component present in the SS amplitude.

In the normalised SS we see an interesting feature develop in all three simulations, there is a local minima and local maxima in $\gamma(k)$. The local minima is caused by the flattening of the PS due to the ionization morphology. The flattening is more prevalent in the PS than the SS, and therefore a minima appears in $\gamma(k)$ at the characteristic scales of ionizing regions. The minima only becomes prevalent during the mid to late stages of reionization once the ionized regions have percolated. The local maxima on the other hand is due to small scale structures. At $\bar{x}_{\rm{HI}}\sim0.75$ the amplitude is tied to the non-linear clustering of the ionized sources and their individual ionized bubbles. This is evident due to the increased amplitude for the Fiducial simulation relative to the \lowMh and \highMh simulations. The Fiducial simulation contains many more smaller mass galaxies increasing the non-linear amplitude on these scales. As reionization proceeds, the maxima grows in amplitude. At $\bar{x}_{\rm{HI}}\sim0.25$ the maxima is significantly larger for the Fiducial simulation relative to the \lowMh and \highMh simulations. The peak is now driven by the prevalence of neutral islands in the IGM (see e.g. Fig \ref{fig:sim-lightcone-slices}). Due to the smaller escape fraction, the Fiducial simulation contains the largest number of neutral islands. On the other hand, the \highMh with its much larger escape fraction ionizes a larger volume per ionizing source, preventing the appearance of neutral islands. Thus the \highMh simulation does not exhibit a strong local maxima.

The location of the local minimum for the Fiducial simulation changes with decreasing neutral fraction from $k\sim0.2\,\rm{Mpc}^{-1}$ to $\sim0.08\,\rm{Mpc}^{-1}$. The characteristic scales of ionizing regions is expected to increase with redshift and decreasing (increasing) neutral (ionization) fraction \citep{Giri2017}. The local minima at $\Bar{x}_{\rm{HI}} \sim 0.5$ appears at $k\sim0.1\,\rm{Mpc}^{-1}$ for all simulations. The expected size of the ionizing regions ranges from $20-100\,\rm{Mpc}$ \citep{Wyithe2004,Zaldarriaga_2004,Lin2016}, with the minima corresponding to size scales from $\sim31-79\,\rm{Mpc}$. The local maxima on the other hand appears at $k\sim1 \,\rm{Mpc}^{-1}$ for the Fiducial, $k\sim0.6 \,\rm{Mpc}^{-1}$ for the \lowMh and $k\sim0.4 \,\rm{Mpc}^{-1}$ for the \highMh simulation. These correspond to sizes $6.3\,\rm{Mpc}$, $3.8\,\rm{Mpc}$, and $1.9\,\rm{Mpc}$ for the Fiducial, \lowMh and the \highMh simulations respectively.  

Fig \ref{fig:PS_SS_Gamma_LX} has the same layout as Fig \ref{fig:PS_SS_Gamma_Mh}, here \lowLX is the double dot dashed line, and the dash dot line is the \highLX simulation. Here the \lowLX simulation power spectra differs from the Fiducial and \highLX at all three ionization states. This difference is driven by the larger absolute temperature difference of the \lowLX simulation compared to the other two simulations. The flattening of the \lowLX PS at $k<1 \,\rm{Mpc}^{-1}$ is still prevalent, since reionization still proceeds the same for the \lowLX, \highLX and Fiducial simulation. The differences at high $k$ are due to the signal temperature on small scales, in the Fiducial and the \highLX simulations the signal is in emission, with the \highLX simulation having a larger emission temperature which causes larger temperature offsets with the ionized regions and thus produces a higher-amplitude 21cm PS. This is not the case for the \lowLX simulation which remains in absorption as reionization progresses.\footnote{It should be noted that at smaller scales $k > 1 \,\rm{Mpc}^{-1}$ approximations in the construction of the simulation yield numerical artefacts that could potentially impact the results.}

When we look at the SS we see a similar trend, with the exception that \lowLX is negative compared to the Fiducial and \highLX scenarios. The shape of \lowLX relative to the other simulations is broadly mirrored about the $k$-axis, and shows a similar morphology to the PS, as is similar with the Fiducial and the \highLX simulations. The normalised SS shows similar features to the Fiducial for both the \lowLX (albeit negative) and for the \highLX. There is a peak/trough at $k\sim0.1 \,\rm{Mpc}^{-1}$, and a subsequent peak/trough at $k\sim1\,\rm{Mpc}^{-1}$. However, this is less pronounced in the \lowLX case, and this feature has more of a flattening from $k = 1-10 \,\rm{Mpc}^{-1}$, which is also seen in the PS. Due to the similarity in the ionization morphology between the Fiducial, \lowLX and the \highLX simulations especially at $x_{\rm{HI}} \sim 0.5$, and $0.25$, the amplitudes of the higher modes ($k>1\,\textrm{Mpc}^{-1})$ appear to be affected more by the X-ray heating, than the ionization morphology itself. 


Fig \ref{fig:PS_SS_Gamma_E0} shows the X-ray threshold simulation PS, SS and normalised SS. Here the double dotted dashed line is the \lowEX simulation, and \highEX is the dashed dotted line. In all cases the SS and the PS of the Fiducial and the \lowEX simulations are almost identical during reionization, differing at most at $k > 1 \,\rm{Mpc}^{-1}$. Again, the ionization morphology for all three simulations is practically identical because they contain the same halo mass distribution and threshold. 

In the case of the \highEX simulation, increasing $E_0$ removes the lowest energy photons, this moves the peak of the X-ray distribution to higher energies. This results in a higher relative fraction of harder X-ray photons. Since harder X-rays heat on larger scales (more uniform heating), this appears to reduce the amplitude on all scales for both the PS and SS respectively. However the reduction seems to impact the SS more than the PS. Fluctuations and non-Gaussianities during reionization are largely driven by ionization morphology at large scales ($k<1\,\textrm{Mpc}^{-1}$) and by the gas density on small scales ($k\geq1\,\textrm{Mpc}^{-1}$) \citep{Wat2018,Maj2018,Ma2023}.

\section{Detectability of the Normalised Skew Spectrum}\label{sec:Detectability}

There is a clear potential advantage to the normalised SS ($\gamma(k)$) compared to the SS. The trough at $k\sim0.1\,\rm{Mpc}^{-1}$ and the peak at $k\sim1\,\rm{Mpc}^{-1}$ demonstrate the sensitivity of the normalised SS to non-Gaussianity in the 21cm signal. Additionally, calculating the SS essentially requires the calculation of the PS, it is therefore straightforward to construct the normalised SS by dividing out the Gaussian amplitude. Naturally the detectability of the features present in $\gamma(k)$, particularly the trough, is important to estimate if there is any potential for its use as a probe of reionization. In this section we investigate the statistical uncertainty of the PS and the SS, where we use these errors to propagate the expected statistical uncertainty on the normalised SS. We save the discussion of system noise from interferometers and foregrounds for future work; these effects deserve independent consideration.

First, we consider the statistical uncertainty of the 21cm PS, also known as the cosmic variance. Therefore of a random Gaussian field, the PS and by extension the variance, describe all the information contained in the field. In this case the uncertainty on the PS in the absence of thermal or instrumental noise is determined by the Poisson sampling:
\begin{equation}\label{eq:Gauss-err}
    \sigma(\Delta^2_{T})(k) = \Delta^2_{T}(k) \sqrt{\frac{(2\pi)^2}{Vk^2\Delta k}}.
\end{equation}

The Poisson sampling error is proportional to one over the square root of the number of Fourier modes for a given spherical shell with width $\Delta k$\footnote{$\Delta k$ is not a constant since logarithmic bins are typically used to calculate the PS. The logarithmic bin width $\log\Delta k = 0.173.$}. The number of modes is proportional to the sampled co-moving volume of space $V$ used to calculate the spectrum. Equation \ref{eq:Gauss-err} assumes Gaussianity, however the signal becomes non-Gaussian as reionization progress \citep{Cooray2005,Furlanetto2006,Wyithe-Morales2007}. These non-Gaussianities correlate the signal on different Fourier modes \citep{cvar-MondalI}, which in turn introduces a noise floor to the expected PS cosmic variance \citep{cvar-MondalII,cvar-MondalIII}:

\begin{equation}\label{eq:Gauss-err+NonGauss}
    \sigma(\Delta^2_{T})(k) = \Delta^2_{T}(k) \sqrt{\frac{(2\pi)^2}{Vk^2\Delta k}} + \sqrt{\frac{T(k,k)}{V}}
\end{equation}

The non-Gaussianities in the PS are proportional to the Trispectrum $T(k,k)$ and $V^{-1/2}$. The effect of these non-Gaussianities is to flatten the signal to noise providing a fundamental limit to the signal detectability \citep{cvar-MondalI}. A calculation of the analytic cosmic variance of the SS is outside of the scope of this work. In this section we assume that the relative uncertainties in the skew-spectrum follow a similar relationship to the PS. Since the SS contains a Gaussian amplitude component, this seems reasonable, and we will demonstrate this in the following sections.

\subsection{Cosmic Variance of the Power Spectrum and the Skew Spectrum}\label{sec:cosmic-var}

\begin{figure*}
    \begin{center}
        \begin{subfigure}[b]{0.495\textwidth}
            \centering
            \includegraphics[width=\textwidth]{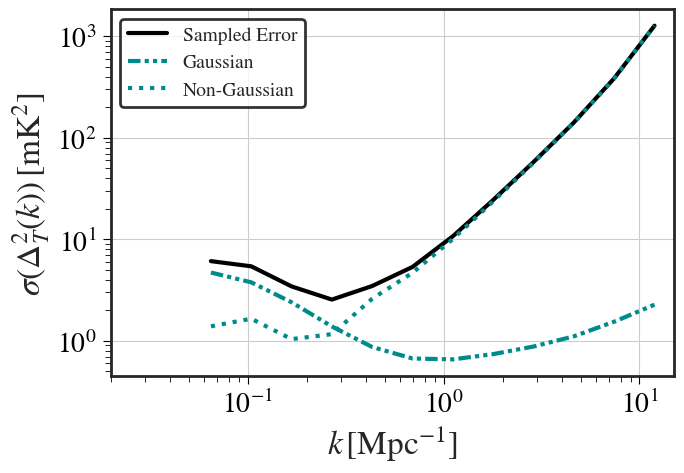}
            \caption[]%
            {{\small $\Delta^2_T (k)$ Uncertainty}}
            \label{fig:u_PS}
        \end{subfigure}
        \hfill
        \begin{subfigure}[b]{0.495\textwidth}
            \centering
            \includegraphics[width=\textwidth]{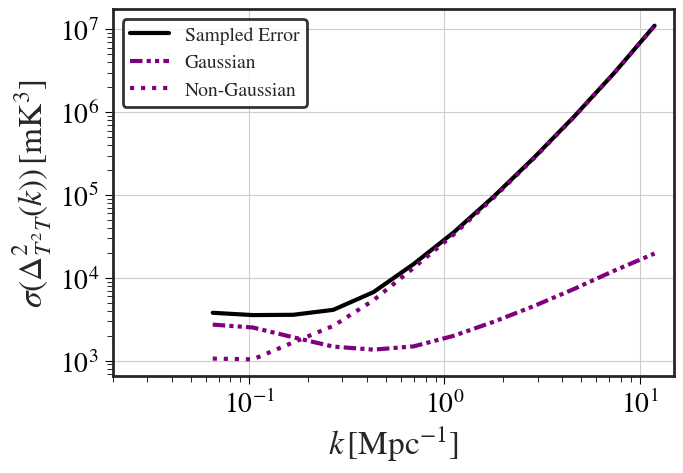}
            \caption[]%
            {{\small $\Delta^2_{T^2T} (k)$ Uncertainty}}
            \label{fig:u_SS}
        \end{subfigure}
        \caption[]
        {\small The numerically estimated statistical uncertainties on PS (a) and the SS (b) (solid line), compared to the expected Gaussian uncertainties (dashed line), and the estimated non-Gaussian component (dotted line).} 
        \label{fig:u_PS_SS}
    \end{center}
\end{figure*}

To investigate the cosmic variance during the EoR, we follow the same method outlined in \citet{Balu_box}. We split each of the simulation volumes in Figs \ref{fig:PS_SS_Gamma_Mh}, \ref{fig:PS_SS_Gamma_LX} and \ref{fig:PS_SS_Gamma_E0} into $27$ sub volumes each with side length $70 \,h^{-1}\, \rm{Mpc}$. We then calculate the PS and SS for each sub volume. The cosmic variance for the PS and the SS is numerically estimated by calculating the variance with respect to the mean power and SS for each $k$-mode. Subfigures \ref{fig:u_PS} and \ref{fig:u_SS} show the sample error (solid black line) for the PS and the SS, compared to the expected Gaussian uncertainty (dashed line) for the Fiducial model at a neutral fraction of $x_{\textrm{HI}} \sim 0.5$. We also estimate the Non-Gaussian component (dotted line) by subtracting the expected Gaussian errors from the sampled errors, these results are similar to those shown in Figure 3 of \citet{Greig_2022c}. We find qualitative agreement with \citet{cvar-MondalII,cvar-MondalIII,Greig_2022c}, where we find an earlier transition to non-Gaussianities in Subfigure \ref{fig:u_PS} in agreement with that seen by \citet{Balu_box}.

The uncertainties at large scales are dominated by the Gaussian component at $(k \lesssim 0.3 \textrm{Mpc}^{-1})$ for the PS and $(k \lesssim 0.2 \textrm{Mpc}^{-1})$ for the SS. This transition is seen at $(k \lesssim 0.5 \textrm{Mpc}^{-1})$ in \citet{cvar-MondalI} and \citet{Greig_2022c}. \citet{Balu_box} attributes the earlier transition to the spin temperature evolution and detailed physical prescriptions for the higher non-Gaussianity in the L210 box. Overall, we find the assumption that the uncertainties in the SS have a similar form as the PS to be a good one.

\subsection{Uncertainty in the Normalised Skew Spectrum}

To estimate the uncertainty in the normalised SS we calculate the linear first order error propagation of equation \ref{eq:[psuedo-skewness]}:
\begin{equation}\label{eq:gamma-var}
    R^2_\gamma(k) = \frac{9}{4} R^2_{T}(k) + R^2_{T^2T}(k) - 3  R_{T}(k)R_{T^2T}(k) \rho(k),
\end{equation}
$(R(k) = \sigma_x(k)/X(k))$ is the relative error for either the PS or the SS, labelled with the subscripts $T$ and $T^2T$ respectively. $\rho(k)$ is the Pearson correlation coefficient of the dimensionless PS and SS as a function of spatial scale. We find significant correlation between the PS and the SS as calculated from the sub volumes in the previous section. 

\begin{figure}
    \centering
    \includegraphics[width=0.475\textwidth]{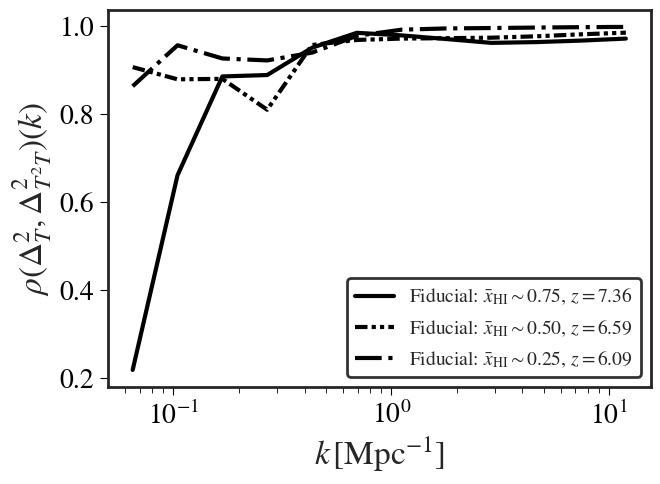}
    \caption{Correlation coefficient between the Fiducial PS and SS for the $x_{\textrm{HI}} \sim 0.75$ (solid), the $0.5$ (dash dotted line), and the $0.25$ (double dot dashed line) neutral fractions.}
    \label{fig:PS_SS_correlation}
\end{figure}

Fig \ref{fig:PS_SS_correlation} shows the correlation for the $\Bar{x}_{\textrm{HI}} \sim 0.75$,\:$0.5$, and the $0.25$ Fiducial simulation coeval boxes as a function of $k$. The average correlation is $0.82$, $0.88$ and $0.92$ respectively for each of the coeval boxes. This result is not surprising and demonstrates that the SS during reionization has significant contribution from the PS amplitude, which is evident from the morphological similarity of the PS and the SS. 

We can derive an expression for the Gaussian component errors in the normalised skew spectrum if we assume that Trispectrum component of $R_{T}$ and $R_{T^2T}$ is zero $(T(k,k)=0)$. In this case the relative errors for the PS and SS are equal $(R_{T}=R_{T^2T})$ and only depend on the volume $V$ and the shell volume $k^2\Delta k$. Equation \ref{eq:gamma-var} therefore simplifies to:

\begin{equation}\label{eq:gamma-var-model}
    R_\gamma(k) = \frac{R_{T}(k)}{2}\sqrt{13 - 12\rho(k)}.
\end{equation}

\begin{figure}
    \centering
    \includegraphics[width=0.475\textwidth]{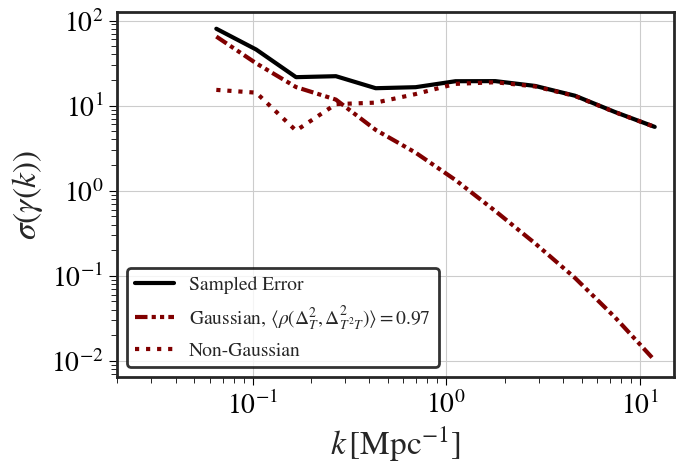}
    \caption{The first order propagated uncertainties on the normalised SS (solid black line), compared to the Gaussian model uncertainties (dashed line), and the estimated non-Gaussian component (dotted line), for the Fiducial simulation at $\Bar{x}_{\textrm{HI}}=0.5$.}
    \label{fig:u_gamma}
\end{figure}

We use Equation \ref{eq:gamma-var-model} as a model for the Gaussian component of the uncertainties in $\gamma(k)$. Fig \ref{fig:u_gamma} shows the sample estimated errors (solid black line) calculated from Equation \ref{eq:gamma-var}. The red double dot dashed line shows the estimated Gaussian uncertainties estimated from Equation \ref{eq:gamma-var-model}. Finally the non-Gaussian component (dotted line) was likewise estimated by subtracting the Gaussian uncertainty model at $\Bar{x}_{\textrm{HI}}=0.5$ from the full uncertainty estimation. The transition from Gaussian dominated to non-Gaussian dominated uncertainties occurs at $k \lesssim 0.3 \textrm{Mpc}^{-1}$. Interestingly, we find that the Non-Gaussian uncertainties are roughly flat as a function of $k$.

\subsubsection{Detection Predictions}

In this section we perform a rudimentary signal to noise ($\textrm{S/N}$) estimate for the normalised SS, for future SKA\_LOW observations. For the estimate we assume the Fiducial simulation as the 21cm signal, and we consider the neutral fractions $\Bar{x}_{\textrm{HI}}=0.25$, $\Bar{x}_{\textrm{HI}}=0.5$, and $\Bar{x}_{\textrm{HI}}=0.75$, and their respective redshifts $7.4$, $6.6$ and $6.1$.

The signal to noise ratio can be defined as the inverse of the relative uncertainty ($\textrm{S/N}(k) = 1/R_\gamma(k)$). In this case we assume the full relative uncertainty for $\gamma(k)$, which includes the non-Gaussianities in Equation \ref{eq:gamma-var}. Notably, $R_\gamma \propto V^{-1/2}$ $(\textrm{S/N} \propto V^{1/2})$ therefore, to estimate the $\textrm{S/N}$ for an SKA\_LOW observation, we can replace the simulation comoving volume ($V^{1/2}_{\rm{sub}}$) by the SKA\_LOW comoving volume ($V^{1/2}_{\rm{SKA}}$). This is done by first calculating the $\textrm{S/N}$ for the Fiducial simulation, then normalising out $V^{1/2}_{\rm{sub}}$, and finally scaling by $V^{1/2}_{\rm{SKA}}$.

To determine the SKA\_LOW comoving volume for an observation, we first need to know the field of view $\Omega_f$ and the observing bandwidth $\Delta\nu$. From these values we can determine the comoving volume for each redshift (neutral fraction) \citep{Hogg2000}. The angular width of the main lobe of the primary beam for an interferometer is approximately given by $\theta\sim\lambda/D$, where $\lambda$ is the observing wavelength, and $D$ is the station diameter which we assume is $35\,\rm{m}$ \citep{SKA}. Using the observing wavelength for each neutral fraction we calculate the field of view to be $2.87^2\,\deg^2$, $2.6^2\,\deg^2$, and $2.44^2\,\deg^2$, for each $\Bar{x}_{\textrm{HI}}$ respectively. Finally assuming an observing bandwidth of $\sim30\,\rm{MHz}$ for each neutral fraction, we then determine the comoving volume using Equation 28 from \citet{Hogg2000}. We then calculate the $\textrm{S/N}$ for each neutral fraction for the Fiducial simulation by taking the ratio of $\gamma(k)/\sigma(\gamma(k))$. The $\textrm{S/N}$ is then scaled by $\sqrt{V_{\rm{SKA}}/V_{\rm{sub}}}$.

\begin{figure}
    \centering
    \includegraphics[width=0.475\textwidth]{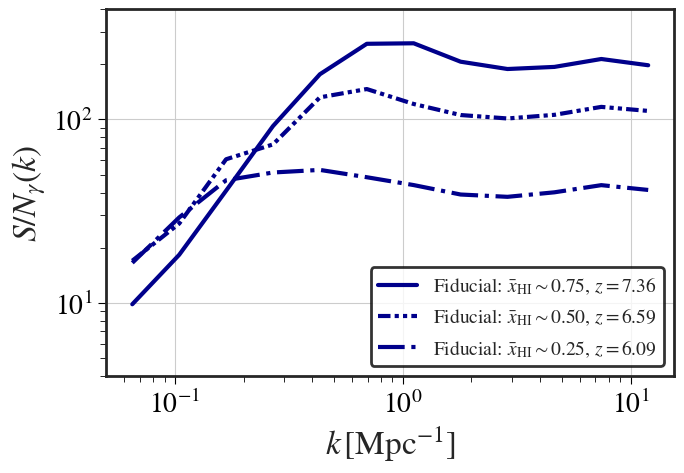}
    \caption{Signal to noise ratio of $\gamma(k)$ for the Fiducial simulation scaled by the expected SKA\_LOW observing comoving volume. The solid curve corresponds to the Fiducial neutral fractions $x_{\textrm{HI}} \sim 0.75$, $0.5$ for the dash dotted line, and $0.25$ for the double dot dashed line.}
    \label{fig:gamma_S2N}
\end{figure}

Fig \ref{fig:gamma_S2N} shows the resulting estimated $\textrm{S/N}$ for $\gamma(k)$ during reionization for $\Bar{x}_{\textrm{HI}}=0.25$ (dashed dot line), $\Bar{x}_{\textrm{HI}}=0.5$ (dash double dot line), and $\Bar{x}_{\textrm{HI}}=0.75$ (solid line). The tapering of the $\textrm{S/N}$ as a function of $k$ is characteristic of the non-Gaussian component in the cosmic variance and is also seen in Figure 4 from \citet{cvar-MondalI}. We see that for all neutral fractions for all spatial scales the $\textrm{S/N} > 10$, with a max signal to noise of $\sim300$ for $\Bar{x}_{\textrm{HI}}=0.75$. Thus, in the absence of thermal noise, foregrounds and systematics, the features present in $\gamma(k)$ should be detectable. MP23 perform their own $\textrm{S/N}$ analysis for the SKA\_LOW with the addition of $1000$h of thermal noise. They find a S/N of $\sim20$ for their SS estimates. In the ideal case assuming foregrounds and systematic noise can be removed, we should expect to be sensitive to the normalised SS trough at $k=0.1\,\textrm{Mpc}^{-1}$ and peak at $k\leq1\,\textrm{Mpc}^{-1}$.

\section{Discussion and Conclusion}\label{sec:Discussion-Conclusion}
%


We investigate the PS and SS as a function of redshift and spatial scale, for a set of seven \MERAXES\:simulations. We vary the halo mass threshold, the X-ray luminosity per star formation rate and the X-ray energy threshold. We find the SS as a function of redshift broadly follows the mean differential temperature brightness as a function of redshift, at large ($k\sim0.1\,\mathrm{Mpc}^{-1}$) and small ($k\sim1\,\mathrm{Mpc}^{-1}$) scales. We do not see a negative sign for the SS amplitude during reionization as seen by MP23; we discuss this in the following subsection~\ref{sec:sign-flip-disc} We further investigate the ionization state and statistics of the IGM during the EoR. We look at the spherically averaged PS, SS and the normalised SS for each simulation set compared to the Fiducial at $\Bar{x}_\textrm{HI}\sim0.75,\:0.5$ and $0.25$. In all simulations we find a local minimum at $k\sim0.1 \,\textrm{Mpc}^{-1}$ during the midpoint of reionization. This minimum corresponds to the characteristic ionization topology (bubble) size during the EoR \citep{Wyithe2004,Bubbles_2,Lin2016}, and we see the evolution of the minima to larger scales with decreasing neutral fraction (increasing ionization fraction). We expect that this feature should be detectable at $k=0.1 \,\textrm{Mpc}^{-1}$ in the absence of instrumental, thermal noise and foreground contamination for the SKA\_LOW and by extension current interferometric experiments. We also see evidence of a local maxima in the halo mass simulations, that shifts to larger scales as a function of the halo mass threshold. This corresponds to small ionized or hot regions around these halos. Our study highlights the importance of higher order statistics, and what additional astrophysical information might be gained from calculating both the SS and the PS.

%
%
The halo mass threshold simulations display the importance of ionization topology as demonstrated in Fig \ref{fig:PS_SS_Gamma_Mh}. This clearly has the biggest impact on the structure present in the normalised SS during reionization. There are however some important limitations and caveats related to the halo mass simulations. These simulations varied the ionization morphology by restricting the halo mass threshold above which galaxies could emit ionizing UV photons, and scaling the amount of UV emission as a function of the halo mass threshold. Galaxies below the threshold however still produced X-ray emission, and thus contribute to heating the IGM. Although this model is nonphysical, it allowed for the halo mass threshold simulations to have a comparable X-ray background to the Fiducial. This effectively isolated the impact of the ionization topology on the 21cm statistics independent of X-ray heating. Changing the X-ray emission in the same manner would delay heating, and have an undesirable impact on the 21cm statistics.

\subsection{Sign of the SS During Reionization}\label{sec:sign-flip-disc}

\begin{figure*}
    \begin{center}
        \begin{subfigure}[b]{0.495\textwidth}
            \centering
            \includegraphics[width=\textwidth]{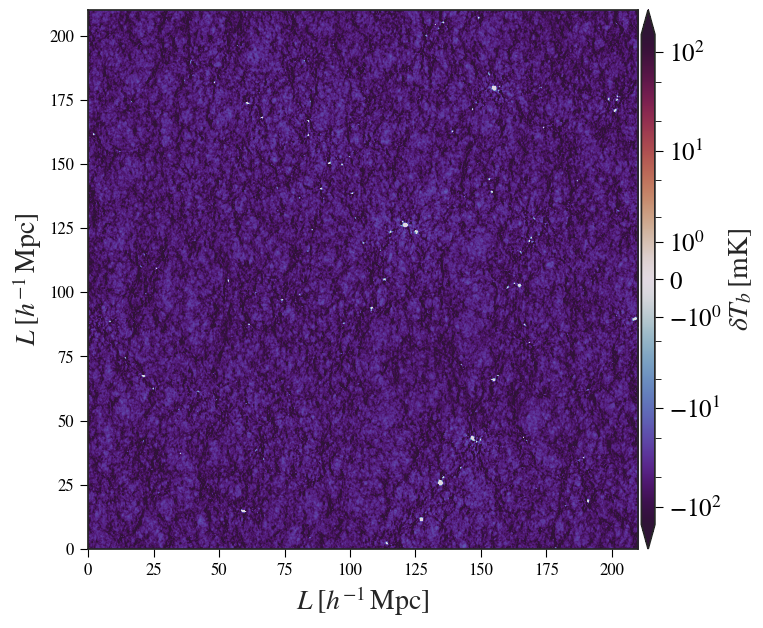}
            \caption[]%
            {{\small \MERAXES\: slice $(x_{\mathrm{HI}}\sim0.98)$}}
            \label{fig:MEREAXES_0pt98}
        \end{subfigure}
        \hfill
        \begin{subfigure}[b]{0.495\textwidth}
            \centering
            \includegraphics[width=\textwidth]{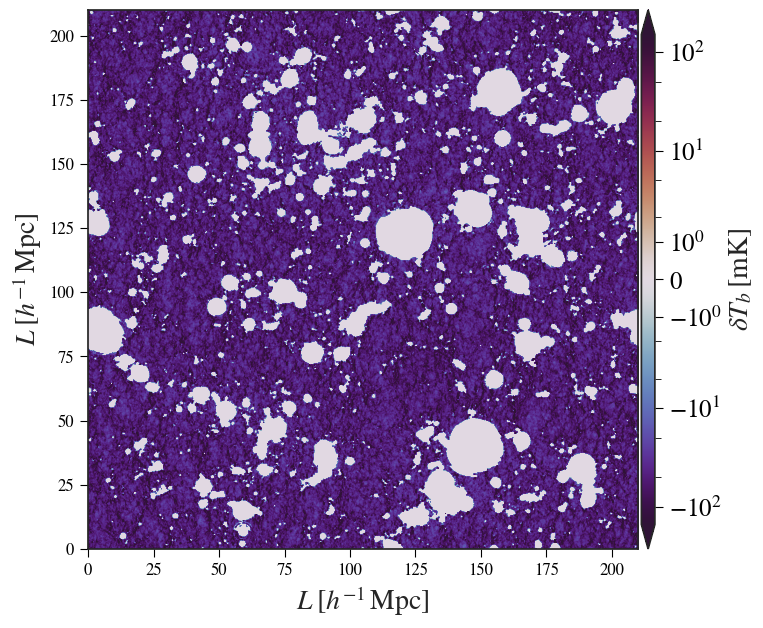}
            \caption[]%
            {{\small \MERAXES\: slice $(x_{\mathrm{HI}}\sim0.75)$}}
            \label{fig:MEREAXES_0pt75}
        \end{subfigure}
        \begin{subfigure}[b]{0.495\textwidth}
            \centering
            \includegraphics[width=\textwidth]{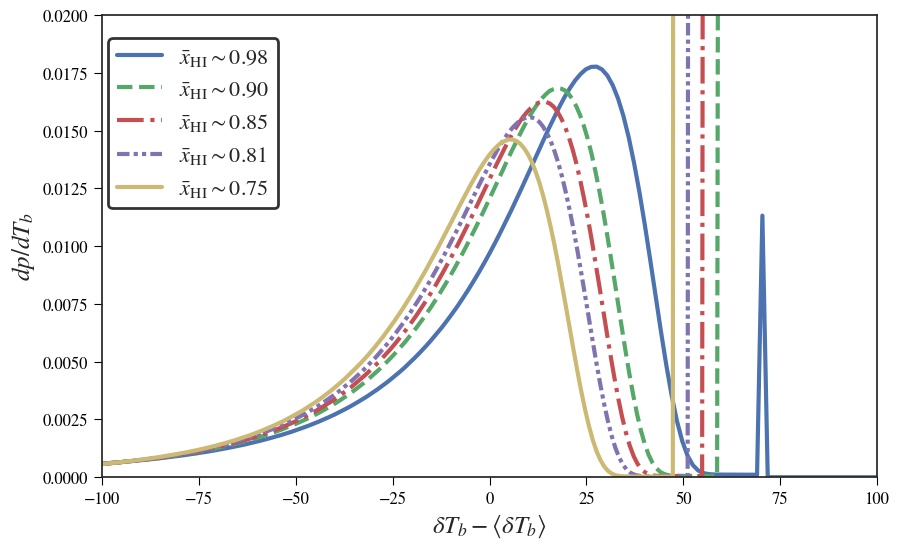}
            \caption[]%
            {{\small \MERAXES\: PDFs}}
            \label{fig:MEREAXES_PDFS}
        \end{subfigure}
        \begin{subfigure}[b]{0.495\textwidth}
            \centering
            \includegraphics[width=\textwidth]{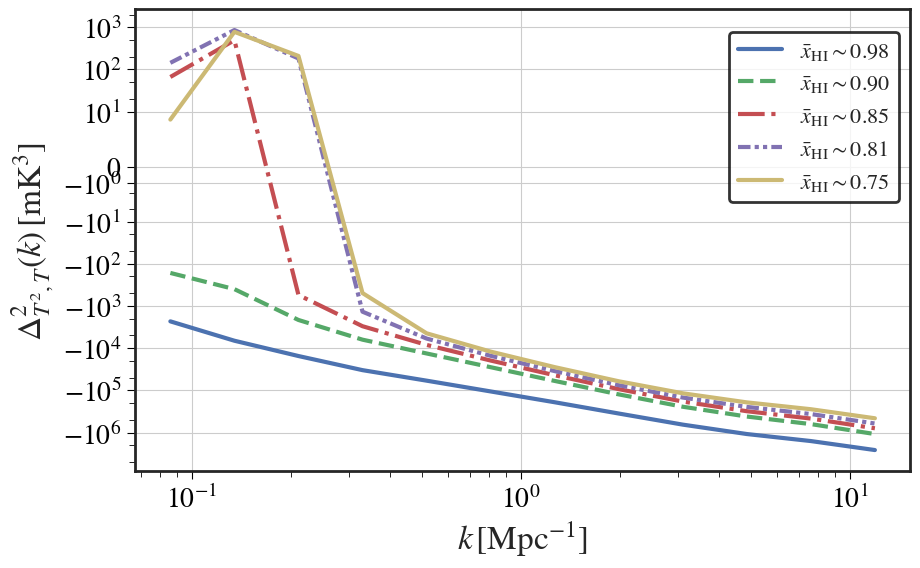}
            \caption[]%
            {{\small \MERAXES\: SS}}
            \label{fig:MEREAXES_SS}
        \end{subfigure}
        \caption[]
        {\small Subfigure (a) shows a 2D slice through the \MERAXES\:$z\sim12$ coeval box, with $x_{\mathrm{HI}}\sim0.98$. Subfigure (b) shows the same slice, with a the $x_{\mathrm{HI}}\sim0.75$ ionization field taken from $z\sim7.24$. Subfigure (c) shows the PDFs for both coeval boxes with the different ionization fields, and subfigure (d) shows the resulting skew spectra for both examples.}
        \label{fig:MERAXES_Tests}
    \end{center}
\end{figure*}
\MERAXES\:like \fastcm calculates the temperature brightness field using the excursion set formalism. However, even though our Fiducial simulation has the same X-ray parameters as the simulations in MP23, we see some significant differences between our results. Most notably the lack of a sign change (positive to negative) during reionization in our SS for all simulations (which include the spin temperature calculation). Additionally, we find the absolute amplitude of our SS is approximately two orders of magnitude greater than MP23, indicating that our differential temperature brightness distribution is significantly more skewed on all scales. We do however see the negative SS amplitude during reionization for the \fidnoTS simulation, similar to \citet{Maj2018}. We note that \citet{Maj2018} perform a similar simulation with \fastcm in the spin saturated limit. However, there are limitations to this assumption and the spin temperature evolution as seen in this work is important for understanding the higher order statistics of the 21cm signal.

The origin of the positive to negative sign transition (or vice versa), is a direct result of the ionization morphology. For a mean zero differential temperature brightness distribution, the ionized cells take the negative value of the mean distribution (prior to mean subtraction). This acts to skew the distribution in the opposite direction at the relevant scales for reionization. The lack of a sign change in \MERAXES\:is due to the significantly larger asymmetry (skewness) in the temperature distribution at all stages of reionization and prior. This larger asymmetry is a direct result of the of the spin temperature evolution in \MERAXES\:compared to \fastcm. A complete comparison of these two simulation packages is far beyond the scope of this paper, here we summarise the key differences which result in the lack of a sign change in our SS during reionization.

Two major difference between \MERAXES\:and \fastcm are the halo mass function (HMF) and the determination of the density fields. The HMF and the density field of \MERAXES\:is determined from the input merger trees of N-body simulations which are inherently non-linear. Additionally, the individual haloes in \MERAXES\:are discrete and treated as independent galaxies which have their own properties. \fastcm on the other hand uses the Press-Schechter conditional HMF renormalised to \citet{SMT2001} , with a continuous density field calculated by smoothing over the perturbed density field. Furthermore, the \citet{SMT2001} HMF is known to overproduce haloes by a factor of two compared to N-body simulations \citep{Watson2013,Murray2013}, thus there are more sources per unit mass when compared to \MERAXES. Another significant difference is how star formation is handled in \MERAXES\: compared to \fastcm. In \MERAXES\: star formation for a galaxy can cease due to feedback effects such as supernovae and AGN, this can lead to intermittent star formation, and thus intermittent heating and UV production. In contrast, in \fastcm, once there is sufficient mass for star formation in a given cell, star formation will continue unabated. Integrating this effect over the time it takes to build the X-ray background, implies that \fastcm produces more sources, and more UV ionization per source compared to \MERAXES. This difference is particularly notable when comparing the depth of the absorption trough for the Fiducial simulation in Fig~\ref{fig:all_sims_meanTb} to the absorption trough in Figure 1 in MP23. The absorption trough for the Fiducial simulation is almost a factor of two deeper than the MP23 simulations, moreover the duration of the heating period is also more than a factor of two greater.

These effects, along with the inherently more skewed non-linear density in \MERAXES\:from the N-body simulations, results in a more negatively skewed temperature brightness distribution at all scales when compared to \fastcm. This asymmetry grows as the IGM adiabatically cools. When the IGM transitions from negative to positive the highly skewed tail also transitions to positive. The inherent asymmetry in the differential temperature distribution outweighs that introduced by the ionization morphology on relevant scales.

The brightness distribution of the simulations in MP23 relative to \MERAXES\:is not as strongly negatively skewed, and thus a small amount of ionization is required to cause the initial sign transition. Therefore, \MERAXES\:with its much larger negative asymmetry requires a significantly larger ionization fraction to cause the equivalent sign flip. To better understand what level of ionization is required to induce a sign transition with \MERAXES, we apply varying levels of ionization fields from lower redshifts to the Fiducial \MERAXES\:simulation at $z\sim12$ ($x_{\mathrm{HI}}\sim0.98$). Specifically, we apply varying levels of ionization morphology from $\Bar{x}_\mathrm{HI}\sim0.98$ to $\Bar{x}_\mathrm{HI}\sim0.75$ to the $z\sim12$ Fiducial simulation volume, to create pseudo-ionization states at $z\sim12$.

A 2D slice of the Fiducial simulation at $z\sim12$ is shown in Subfig~\ref{fig:MEREAXES_0pt98}, and in Subfig~\ref{fig:MEREAXES_0pt75} we show the same slice with the $\Bar{x}_\mathrm{HI}\sim0.75$ ($z\sim7.24$) ionization field applied. We calculate the differential temperature brightness distribution for each pseudo-ionization state, these are displayed in Subfig ~\ref{fig:MEREAXES_PDFS}, with the original box in solid blue. The decreasing ionization fraction results in shifting the peak of the distribution closer to zero. We then calculate the SS for each pseudo-ionization state, these are shown in Subfig~\ref{fig:MEREAXES_SS}. At large scales ($k\sim0.1\,\mathrm{Mpc}^{-1}$) for ionization states of $\Bar{x}_\mathrm{HI}<0.9$ (greater than $10\%$ ionization) the SS is positive, compared to the initial negative SS of the Fiducial simulation at $z\sim12$. This is similar to the results seen in Figure 3 of MP23.
%

\subsection{Future Work}
%
%
Future work should consider how the ionization topology is quantitatively linked to the trough feature seen in the halo mass simulations. This could be investigated by measuring the characteristic scales of ionization for the coeval boxes in this work using the methods outlined in \citet{Lin2016}, in particular the mean free path method. Understanding the characteristic ionization scale and variance, and comparing them to the trough position and width are important for quantitatively understanding how the ionization topology affects the higher order statistics. This could additionally be performed on a simple toy model similar to the one used in \citet{Maj2018}, where the characteristic scales can be varied to understand their impact on the observed features in the normalised SS, independent of other affects such as X-ray heating.


This work highlights the importance of the spin temperature in the higher order statistics of the expected 21cm signal. Furthermore, this work demonstrates the impact of the different implementations of star formation in the 21cm signal between \MERAXES\:and \fastcm. More work is required to understand the significance of these differences and how they impact the expected higher order statistics of the 21cm signal.

Future work will consider the practicality of calculating the SS with the current and future generations of radio interferometric instruments. The squared sky temperature can not be measured directly, so we must estimate it from interferometric visibilities derived from observations\footnote{In Section \ref{sec:estimating-SS} of the appendix, we outline one method and its challenges for estimating the SS from radio interferometric visibilities.}. This ultimately involves a convolution of the signal in Fourier space or a multiplication in image space and subsequent inversion back to Fourier space. Each additional step in the process propagates systematic and instrumental effects and spreads them across different Fourier modes. These effects already impact the PS for numerous experiments which are systematics limited rather than thermal noise limited. Further investigation is needed to understand how these effects propagate through the SS, and whether they render a realistic measurement impractical. 

\section*{Acknowledgements}
We acknowledge the helpful discussion of Simon Mutch who assisted with this work. This research was supported by the Australian Research Council Centre of Excellence for All Sky Astrophysics in 3 Dimensions (ASTRO 3D) through project number CE170100013. JHC is supported by a Research Training Program scholarship. CMT is supported by an ARC Future Fellowship under grant FT180100321.
The International Centre for Radio Astronomy Research (ICRAR) is a Joint Venture of Curtin University and The University of Western Australia, funded by the Western Australian State government. 
The MWA Phase II upgrade project was supported by the Australian Research Council LIEF grant LE160100031 and the Dunlap Institute for Astronomy and Astrophysics at the University of Toronto.
This scientific work makes use of the Murchison Radio-astronomy Observatory, operated by CSIRO. We acknowledge the Wajarri Yamatji people as the traditional owners of the Observatory site. Support for the operation of the MWA is provided by the Australian Government (NCRIS), under a contract to Curtin University administered by Astronomy Australia Limited. We acknowledge the Pawsey Supercomputing Centre which is supported by the Western Australian and Australian Governments.

Part of this work was performed on the OzSTAR national facility at the Swinburne University of Technology. The OzSTAR program partially receives funding from the Astronomy National Collaborative Research Infrastructure Strategy (NCRIS) allocation provided by the Australian Government. This research was also undertaken with the assistance of resources from the National Computational Infrastructure (NCI Australia), an NCRIS-enabled capability supported by the Australian Government. 

\section*{Data Availability}

The simulated data referenced in this work, is available upon reasonable request to the primary author.



\bibliographystyle{mnras}
\bibliography{main} 




\appendix

\section{Estimating the SS From Visibilities}\label{sec:estimating-SS}
The most straightforward method for estimating the SS, is to calculate it from image lightcones constructed form radio interferometric visibilities. In this case, visibilities measured at different frequencies are used to constructed image slices at each frequency. These images can then be mean subtracted, squared and Fourier transformed. Multiplying the quadratic Fourier temperature field by the conjugate of the Fourier temperature field, and averaging spherically will estimate the SS. There are however numerous challenges. First assume an ideal scenario where the foregrounds are perfectly subtracted, and are not thermal noise limited. The image noise may not be entirely Gaussian due to the spatial correlations introduced by the image PSF. Additionally, squaring the image changes the noise statistics. Furthermore, primary beam correcting the image would upscale the image noise. A solution might be to consider smaller image subset where the primary beam is relatively constant. This will limit the Fourier space resolution, but may be a desirable trade off when considering instrumental effects introduced by the primary beam and the image noise. This could be achieved with a spatial taper such as a circular top hat to limit aliasing in Fourier space. The other limitation to this method is the computational cost of imaging each channel, for low frequency radio interferometers there are typically of order $100$ channels for a full bandwidth observation. One solution could be to image only a subset of channels, or to  average channels together at the expense of bandwidth smearing effects \citep{Bridle}. Estimating the SS from radio interferometric measurements, and the associated challenges will be focus of future work.

\bsp	
\label{lastpage}
\end{document}